\documentclass[twocolumn,aps,showpacs,showkeys,prb
   ,superscriptaddress,reprint]{revtex4-1}
\usepackage{natbib}
\usepackage{url}
\usepackage{bm}
\usepackage{textcase}
\usepackage{color}
\usepackage{graphicx}
\usepackage{amsmath}
\usepackage{ulem}
\usepackage[colorlinks=true,citecolor=blue]{hyperref} 

\hypersetup{pdftitle={Reduced density-matrix functionals applied to
    the Hubbard dimer}}

\hypersetup{pdfauthor={E. Kamil, R. Schade, Th. Pruschke, P.E. Bloechl}}
\hypersetup{pdfdisplaydoctitle}
\DeclareMathOperator*{\stat}{stat}
\newcommand{\mat}[1]{{\boldsymbol #1}}
\newcommand{\e}[1]{{\rm e}^{#1}}
\newcommand{\Tr}{{\rm Tr}}

\begin{document}
\title{Reduced density-matrix functionals applied to the Hubbard dimer}

\author{Ebad Kamil}

\affiliation{Institute for Theoretical Physics,
  Georg-August-Universit\"at G\"ottingen, Friedrich-Hund-Platz 1,
  37077 G\"ottingen, Germany} 

\author{Robert Schade}

\affiliation{Institute for Theoretical Physics, Clausthal University
  of Technology, Leibnizstr. 10, 38678 Clausthal-Zellerfeld, Germany}

\author{Thomas Pruschke}

\affiliation{Institute for Theoretical Physics,
  Georg-August-Universit\"at G\"ottingen, Friedrich-Hund-Platz 1,
  37077 G\"ottingen, Germany} 

\author{Peter E. Bl\"ochl}
\affiliation{Institute for Theoretical Physics, Clausthal University
  of Technology, Leibnizstr. 10, 38678 Clausthal-Zellerfeld, Germany}

\affiliation{Institute for Materials Physics,
  Georg-August-Universit\"at G\"ottingen, Friedrich-Hund-Platz 1,
  37077 G\"ottingen, Germany}

\date {\today}
 
\begin{abstract}
Common density-matrix functionals, the M\"uller and the power
functional, have been benchmarked for the half-filled Hubbard dimer,
which allows to model the bond dissociation problem and the transition
from the weakly to the strongly correlated limit. Unbiased numerical
calculations are combined with analytical results.  Despite the well
known successes of the M\"uller functional, the ground state is
degenerate with a one-dimensional manifold of ferromagnetic solutions.
The resulting infinite magnetic susceptibility indicates another
qualitative flaw of the M\"uller functional. The derivative
discontinuity with respect to particle number is not present
indicating an incorrect metal-like behavior. The power functional
actually favors the ferromagnetic state for weak interaction.
Analogous to the Hartree-Fock approximation, the power functional
undergoes a transition beyond a critical interaction strength, in this
case however, to a non-collinear antiferromagnetic state.
\end{abstract}
\maketitle
 \section{Introduction}
Ab-initio calculations are dominated by density functional theory
(DFT)\cite{hohenberg64_pr136_B864,kohn65_pr140_1133}, which provides
an efficient and accurate description of the electronic structure for
most materials\cite{cramer09_pccp11_10757}. For materials with strong
correlations, however, many of the available density functionals yield
poor results \cite{vonbarth04_physicascripta109_9,cohen12_cr112_289}.
Most well known is the case of transition metal oxides, for which most
density functionals produce a qualitatively incorrect
description\cite{terakura84_prb30_4734}.
However, also elementary chemical processes such as bond dissociation
are described poorly by currently available density
functionals.\cite{cohen12_cr112_289}

There is a quest to improve the description by borrowing from methods
specifically designed for strongly correlated materials. Among them
are LDA+U\cite{anisimov91_prb44_943}, DFT-plus-dynamical
mean-field theory
\cite{georges96_rmp68_13,held07_advp56_829,kotliar06_rmp78_865},
and DFT-plus-Gutzwiller approximation\cite{buenemann98_prb57_6896,
  schickling14_njp16_83034,wang08_prl101_66403}. The guiding idea
behind these approaches is to merge density functional theory with
methods developed for the study of strong correlations for model
Hamiltonians such as the Hubbard model.\cite{gutzwiller63_prl10_159,
  hubbard63_prsla276_238, kanamori63_progtheorphys30_275}

We consider reduced density-matrix functional theory
(rDMFT)\cite{gilbert75_prb12_2111,levy79_pnas76_6062} to be a useful
framework for a rigorous formulation of such hybrid
theories.\cite{bloechl11_prb84_205101,bloechl13_prb88_25139} Reduced
density-matrix functional theory can be viewed as a relative of DFT,
which emphasizes orbital occupations rather than the density as basic
variable. Such a description seems to be natural for correlated
materials, because the latter are dominated by orbital physics.

The link from rDMFT to many-particle wave functions has been
established by Levy's constrained-search
algorithm\cite{levy79_pnas76_6062} on the one hand.
The link to many-body perturbation theory and
Green's function, on the other hand, has been provided
recently\cite{bloechl13_prb88_25139} via the Luttinger-Ward
functional\cite{luttinger60_pr118_1417}.

In order to avoid the full complexity of an explicit many-body
description, most density-matrix functionals are not extracted from
the exact expressions\cite{levy79_pnas76_6062,bloechl13_prb88_25139}.
Rather, one proceeds analogously to the development of density
functionals, namely by searching models\cite{mueller84_pl105A_446,
  goedecker98_prl81_866, gritsenko05_jcp122_204102,
  sharma08_prb78_201103, marques08_pra77_32509} for the density-matrix
functional, that capture the most essential physical effects while
having an algebraic dependence on the density matrix.

The development of such model density-matrix functionals relies on
benchmark systems that allow one to evaluate their quality. Of
particular interest are exactly solvable problems. Such studies have
been performed for the Moshinsky
atom\cite{benavidesriveros12_epjd66_274}, the homogeneous electron
gas\cite{lathiotakis09_pra79_40501} and the Hubbard
model\cite{olsen14_jcp140_164116,disabatino15_jcp143_24108}. Di
  Sabatino et al.\cite{disabatino15_jcp143_24108} 
performed an in-depth
  analysis of the method proposed by Sharma et
  al.\cite{PhysRevLett.110.116403} to evaluate the spectral function
  of the Hubbard dimer from the M\"uller density-matrix
  functional\cite{mueller84_pl105A_446}.

As pointed out by Cohen, Sanchez and Yang\cite{cohen12_cr112_289},
many of the failures of current density functionals for correlated
materials can be traced back to the derivative discontinuities present
in a surprisingly simple system, namely the hydrogen or helium dimer
in different charge states, i.e. H$_2^+$, H$_2$, He$_2^+$. Therefore,
the two-site Hubbard model, the Hubbard dimer, can be considered as
model system for the correlation effects present in a chemical bond.

The most prominent failure of density functionals occurs during bond
dissociation. If we denote the hopping parameter between the bonded
atoms with $t$ and the on-site interaction strength with $U$, bond
dissociation is described by the limit $t\rightarrow 0$ at constant
$U$. Thus, the system evolves from a weakly correlated state into the
strongly correlated limit $U/t\rightarrow\infty$ as the bond is
broken. The
  large-interaction limit $U\rightarrow\infty$ of the Hubbard
  model differs from the bond-dissociation limit only by the choice of
  the energy scale.

One of the major arguments in favor of density-matrix functionals is
that one of the most simple functionals, the M\"uller
functional\cite{mueller84_pl105A_446}, seems to provide a correct
description of the bond-dissociation problem, for which common
density functionals fail\cite{0953-8984-27-39-393001}.  In this
paper we study the performance of a class of commonly used model
density-matrix functionals for the half-filled Hubbard dimer. We point
out that, despite some successes, also these density-matrix
functionals reproduce a number of features in a qualitatively
incorrect manner. Thus, this work sets the stage for the development
of entirely new class of functionals.\cite{bloechl13_prb88_25139}

In section~\ref{sec:theory}, we define our notation and introduce the
basic concepts of density-matrix functionals. In
section~\ref{sec:hubbarddimer}, we present the analytically exact
treatment of the Hubbard dimer and describe its relevant properties.
In section~\ref{sec:methods}, we describe the numerical methodology of
searching for the ground state for the model density-matrix
functionals. In section~\ref{sec:performance}, we describe the results
obtained with the Hartree-Fock approximation and the commonly used
M\"uller and power functionals. In section~\ref{sec:beyondhalffilling}
we study the Hubbard dimer beyond the half filling and in
section~\ref{sec:larger systems} we discuss briefly the
transferability of our results to larger systems.  Our results are
summarized in section~\ref{sec:conclusion}.

\section{Theoretical framework}
\label{sec:theory}
\subsection{General many-particle problem}
The many-particle Hamiltonian for interacting electrons can be
expressed in terms of field operators $\hat{\psi}(\vec{x})$ and
$\hat{\psi}^\dagger(\vec{x})$ in the form
\begin{eqnarray}
\hat{H}&=&\int d^4x\;\hat{\psi}^\dagger(\vec{x})
\left(\frac{-\hbar^2}{2m_e}\vec{\nabla}^2+v_{ext}(\vec{x})\right)
\hat{\psi}(\vec{x})
\nonumber\\
&+&\frac{1}{2}\int d^4x\int d^4x'\;
\hat{\psi}^\dagger(\vec{x})\hat{\psi}^\dagger(\vec{x'})
\frac{e^2}{4\pi\epsilon_0|\vec{r}-\vec{r'}|}
\hat{\psi}(\vec{x'})\hat{\psi}(\vec{x}),
\nonumber\\
\end{eqnarray}
where $\vec{x}=(\vec{r},\sigma)$ is a combined position and spin
variable. We use the shorthand $\int d^4x=\sum_\sigma\int d^3x$ for
the integration over positions and the sum over spin indices.
The field operators obey the usual anticommutator relations
$\left[\hat{\psi}^\dagger(\vec{x}),\hat{\psi}(\vec{x'})\right]_+
=\delta(\vec{r}-\vec{r'})\delta_{\sigma,\sigma'}$.

A discrete, orthonormal one-particle basis set
$\{\chi_\alpha(\vec{x})\}$ determines the creation and annihilation
operators of electrons in the one-particle orbitals
\begin{eqnarray}
\hat{c}^\dagger_\alpha&=&
\int d^4x\; \chi_\alpha(\vec{x})\hat{\psi}^\dagger(\vec{x})
\nonumber\\
\hat{c}_\alpha&=&
\int d^4x\; \chi^*_\alpha(\vec{x})\hat{\psi}(\vec{x}).
\end{eqnarray}

In this one-particle basis set we obtain the discrete Hamiltonian
 \begin{equation}
 \label{eq:Hamiltonian}
 \mathcal{H}=\sum_{\alpha\beta}h_{\alpha\beta} 
\hat{c}_{\alpha}^{\dagger}\hat{c}_{\beta}^{\phantom\dagger}
+\frac{1}{2}\sum_{\alpha\beta\gamma\delta}U_{\alpha\beta\gamma\delta}\,
\hat{c}_{\alpha}^{\dagger}\hat{c}_{\beta}^{\dagger}
\hat{c}_{\delta}^{\phantom{\dagger}}\hat{c}_{\gamma}^{\phantom{\dagger}}
\end{equation}
 with the one-particle Hamiltonian
\begin{eqnarray}
h_{\alpha,\beta}=\int d^4x\;\chi^*_a(\vec{x})
\left(\frac{-\hbar^2}{2m_e}\vec{\nabla}^2+v_{ext}(\vec{x})\right)
\chi_\beta(\vec{x}).
\end{eqnarray}
The off-diagonal elements of $\mat{h}$ are named hopping parameters,
and the diagonal elements are named orbital energies.

The interaction matrix elements are
\begin{equation}
  \label{eq:W_tensor}
  U_{\alpha\beta\gamma\delta}=
\int d^4x\int d^4x'\; 
\frac{e^2 
\chi^*_{\alpha}(\vec{x})
\chi^*_{\beta}(\vec{x'})
\chi_{\gamma}(\vec{x})
\chi_{\delta}(\vec{x'})}
{4\pi\epsilon_0|\vec r-\vec r'|}.
\end{equation}

\subsection{One-particle reduced density matrix}
\label{sec:foundation_denmat}
The one-particle reduced density matrix of an ensemble of fermionic
many-particle wave functions $|\Phi_{j}\rangle$ with probabilities
$P_{j}$ is defined as
 \begin{equation}
  \rho_{\alpha\beta}
=\sum_j P_j\langle\Phi_j|\hat{c}_{\beta}^{\dagger}
\hat{c}_{\alpha}^{\phantom\dagger}|\Phi_j\rangle.
 \label{eq:One_particleDM1}
 \end{equation}

The density matrix is often represented by the eigenvalues and
eigenstates of the corresponding one-particle operator 
 \begin{equation}
 \label{eq:One_particleDM2}
  \hat{\rho}=\sum_{\alpha\beta}|\chi_\alpha\rangle\rho_{\alpha,\beta}
\langle\chi_\beta|.
 \end{equation}
 The eigenvalues of $\hat{\rho}$ are the occupations $f_{n}$ and the
 eigenstates $|\phi_n\rangle$ are named natural
 orbitals\cite{loewdin55_pr97_1474}. Thus the density matrix can be
 expressed by its eigenvalues and eigenstates in the form
\begin{eqnarray}
\rho_{\alpha,\beta}
=\sum_n\langle\chi_\alpha|\phi_n\rangle f_n\langle\phi_n|\chi_\beta\rangle.
 \label{eq:One_particleDMnatorb}
\end{eqnarray}

Not every hermitian matrix can be also be represented as the
one-particle reduced density matrix of an ensemble of many-particle
wave functions according to Eq.~\eqref{eq:One_particleDM1}.  A matrix
that can be represented by an ensemble of fermionic N-particle wave
functions is called ensemble N-representable.
Coleman\cite{coleman63_rmp35_668} has shown that eigenvalues of all
ensemble N-representable one-particle reduced density matrices lie
between zero and one and that all hermitian matrices with eigenvalues
between zero and one are ensemble N-representable.

\subsection{Helmholtz potential and density-matrix functional}
The Helmholtz potential $A_{\beta,\mu}[\mathcal{H}]$, the
thermodynamic potential for finite temperature and fixed particle
number, for a many-particle system can be expressed with the help of
the density-matrix functional $F^{\hat{W}}_\beta[\mat{\rho}]$
as\cite{levy79_pnas76_6062,lieb83_ijqc24_243,
  PhysRevA.92.052514,bloechl13_prb88_25139}
 \begin{widetext}
\begin{eqnarray}
\label{eq:Omega}
A_{\beta,N}[\hat{h}+\hat{W}]
&=&\min_{|\phi_n\rangle,f_n\in[0,1]}\stat_{\Lambda,\mu}
\biggl\lbrace\sum_n f_n\langle\phi_n|\hat{h}|\phi_n\rangle
+F_\beta^{\hat{W}}\Bigl[\sum_n|\phi_n\rangle f_n\langle\phi_n|\Bigr]
\nonumber\\
&-&\mu\left(\sum_n f_n-N\right)
-\sum_{m,n}\Lambda_{m,n}\Bigl(\langle\phi_n|\phi_m\rangle-\delta_{m,n}\Bigr)
\biggr\rbrace
\nonumber\\
\label{eq:gcpdm2}
\end{eqnarray}
 \end{widetext}
where $\hat{h}=\sum_{\alpha,\beta}|\chi_\alpha\rangle
h_{\alpha,\beta}\langle\chi_\beta|$.

The reduced density-matrix functional $F_\beta^{\hat{W}}[\mat{\rho}]$
is  universal in the sense that it depends only on the intrinsic
properties of the electron gas, namely the interaction $\hat{W}$,
while it is independent of the one-particle Hamiltonian $\hat{h}$.
The chemical potential $\mu$ is a Lagrange multiplier that constrains
the electron number to $N$. $\Lambda_{mn}$ are the Lagrange
multipliers which enforce that natural orbitals $|\phi_m\rangle$
remain orthonormal.

The reduced density-matrix functional 
\begin{eqnarray}
F^{\hat{W}}_{\beta}[\mat{\rho}]=
E_H[\mat{\rho}]+U_{xc,\beta}[\mat{\rho}]
\end{eqnarray}
is the sum of Hartree energy $E_H$ and the exchange-correlation energy
$U_{xc}$

The Hartree energy $E_{H}[\mat{\rho}]$ is obtained from the
electron density
\begin{eqnarray}
n(\vec{r})=\sum_\sigma\sum_{\alpha,\beta}
\chi_\alpha(\vec{r},\sigma)\rho_{\alpha,\beta}
\chi^*_\beta(\vec{r},\sigma)
\end{eqnarray}
as
\begin{eqnarray}
E_{H}[\mat{\rho}]&=&\frac{1}{2}\int d^3r\int d^3r'\;
\frac{e^2n(\vec{r})n(\vec{r'})}{4\pi\epsilon_0|\vec{r}-\vec{r'}|}
\nonumber\\
&=&\frac{1}{2}\sum_{\alpha,\beta,\gamma,\delta}U_{\alpha,\beta,\delta,\gamma}
\rho_{\delta,\alpha}\rho_{\gamma,\beta}.
\label{eq:ehartree}
\end{eqnarray}

The exchange-correlation energy $U_{xc}$ contains the complexity of
the many particle problem.  It is the electrostatic interaction of
each electron with its exchange-correlation hole and the entropy term
$-TS$. It should be noted that the exchange-correlation energy
$E_{xc}$ of DFT also contains a contribution from the kinetic energy,
which is absent in the quantity $U_{xc}$ used in rDMFT.

We restrict the present study to zero temperature and thus ignore the
entropy term. To keep the notation simple, we suppress the index for
the inverse temperature in the remainder of the text.

Having laid down the basic concepts and our notation, we proceed with
the concept of the hole function as a tool for the construction of
approximate density-matrix functionals.

\subsection{Hole function and the construction of 
density-matrix functionals}
 \subsubsection{Hole function}
In this section we discuss several exact properties of the hole
function, which have been central to the development of density
functionals, and in the following section we outline its role for the
construction of approximate density-matrix functionals.

The hole function $h(\vec{r},\vec{r'})$
allows to express the two-particle density $n^{(2)}(\vec{r},\vec{r'})$
in the form
\begin{eqnarray}
n^{(2)}(\vec{r},\vec{r'})=n(\vec{r})n(\vec{r'})+n(\vec{r})h(\vec{r},\vec{r'}).
\end{eqnarray}
Note that  the interaction-strength
averaged hole function is used in DFT, while in rDMFT, the hole function at
full interaction strength is of interest.

The hole function integrates to minus one,
\begin{equation}
\int d^3r'\; h(\vec{r},\vec{r'})=-1\;,
\label{eq:sumrule}
\end{equation}
and it is always negative.\cite{perdew96_prb54_16533}

 These conditions constrain the shape of the hole function strongly,
 so that the exchange-correlation energy can be predicted reasonably
 well already with simple assumptions about the hole function.  An
 insightful description of the hole function, which guided the
 development of a number of density-matrix functionals, has been given
 by Baerends and
 Buijse\cite{baerends01_prl87_133004,buijse02_molphys100_401}.

In the Hartree-Fock approximation, the hole function has the form
\begin{eqnarray}
h(\vec{r},\vec{r'})=\frac{-1}{n(\vec{r})}\sum_{m,n}f_mf_n
\sum_{\sigma,\sigma'}
\phi^*_m(\vec{x})\phi_n(\vec{x})
\phi^*_n(\vec{x'})\phi_m(\vec{x'}).
\nonumber\\
\label{eq:hfholefunction}
\end{eqnarray}
As a consequence of the orthonormality of the natural orbitals, the sum
rule Eq.~\eqref{eq:sumrule} is obtained as
\begin{equation}
  \int d^3r'\;h(\vec{r},\vec{r'})=
\frac{-1}{n(\vec{r})}\sum_{n\sigma}f_n^2 \phi^*_n(\vec{x})\phi_n(\vec{x})=-1.
\end{equation}
The sum-rule is fulfilled exactly, when $f_n^2=f_n$ that is for
integer occupations. For fractional occupations, however, the
Hartree-Fock expression violates the sum-rule.

The exchange-correlation term in the Hartree-Fock approximation is
\begin{eqnarray}
U_{xc}^{HF}[\mat{\rho}]
&=&
-\frac{1}{2}\sum_{m,n}f_{m}f_{n} 
\sum_{\alpha\beta\gamma\delta}U_{\alpha\beta,\delta\gamma}
\nonumber\\
&&\times\langle\chi_{\gamma}|\phi_{m}\rangle
\langle\phi_{m}|\chi_{\alpha}\rangle
\langle\chi_{\delta}|\phi_{n}\rangle
\langle\phi_{n}|\chi_{\beta}\rangle.
\label{eq:HFK}
\end{eqnarray}

\subsubsection{Construction of density-matrix functionals}
\label{sec:construction_of_density_matrix_functionals}
Most empirical density-matrix functionals maintain this general form
of the Hartree-Fock exchange term, 
\begin{eqnarray}
\label{eq:cmn}
U_{xc}[\mat{\rho}]
&=& 
-\frac{1}{2}\sum_{m,n}c_{m,n} 
\sum_{\alpha\beta\gamma\delta}U_{\alpha\beta,\delta\gamma}
\nonumber\\
&&\times\langle\chi_{\gamma}|\phi_{m}\rangle
\langle\phi_{m}|\chi_{\alpha}\rangle
\langle\chi_{\delta}|\phi_{n}\rangle
\langle\phi_{n}|\chi_{\beta}\rangle,
\end{eqnarray}
but replace the factor $f_nf_m$ in Eq.~\eqref{eq:HFK}
by coefficients $c_{m,n}$ with a different dependence on the occupations.

Taking the hole-function in the Hartree-Fock approximation
Eq.~\eqref{eq:hfholefunction} as a starting point,
M\"uller\cite{mueller84_pl105A_446} has shown that one can enforce the
sum rule Eq.~\eqref{eq:sumrule} also for fractional occupations with
an ansatz
\begin{equation}
h(\vec{r},\vec{r'})=\frac{1}{n(\vec{r})}\sum_{m,n}c_{n,m}^M
\sum_{\sigma,\sigma'}
\phi^*_m(\vec{x})\phi_n(\vec{x})
\phi^*_n(\vec{x'})\phi_m(\vec{x'})
\label{eq:construction_rdmf_ansatz_mueller}
\end{equation}
with $c^M_{m,n}=\frac{1}{2}(f_m^{\frac{1}{2}+p}f_n^{\frac{1}{2}-p}
+f_m^{\frac{1}{2}-p}f_n^{\frac{1}{2}+p})$.

M\"uller identified $p=0$ as the choice that minimizes the violation
of the positive definiteness of the hole function. This is the value
used in nearly all applications.

Later, Sharma et al.\cite{sharma08_prb78_201103} invented the so-called
power functional by introducing an additional parameter $\alpha$ in
the dependence of the coefficients $c_{m,n}$ on the occupations. They
chose the form $c_{m,n}^P(\alpha)=f_m^\alpha f_n^\alpha$ that smoothly
interpolates between the M\"uller functional with $\alpha=1/2$ and the
Hartree-Fock approximation with $\alpha=1$. The main reason for this
construction is according to Sharma et al.\cite{sharma08_prb78_201103}
the well known overcorrelating behavior of the M\"uller functional,
that will be mediated by a parameter $\alpha>1/2$.  The coefficients
$c_{m,n}$ for the approximate density-matrix functionals considered in
this work are summarized in table~\ref{tab:parms}.
\begin{table}[h!]
{\renewcommand{\arraystretch}{1.5}
\begin{tabular}{|l|l|}
\hline
\hline
Hartree-Fock approximation & $c^{HF}_{m,n}=f_mf_n$\\
M\"uller functional\cite{mueller84_pl105A_446} 
& $c^{M}_{m,n}=f_m^\frac{1}{2}f_n^\frac{1}{2}$\\
power functional\cite{sharma08_prb78_201103} 
& $c^{P}_{m,n}(\alpha)=f_m^\alpha f_n^\alpha$\\
\hline
\hline
\end{tabular}
}
\caption{\label{tab:parms}Dependence of the parameters $c_{m,n}$ on
  the occupations $f_n$ as defined in Eq.~\eqref{eq:cmn} for
  density-matrix functionals used in this work.  }
\end{table}

\section{Hubbard dimer}
\label{sec:hubbarddimer}
The two-site Hubbard model, the Hubbard dimer, is the simplest model
for the covalent bond and bond breaking. 

The Hubbard dimer has a one-particle basis with four spin orbitals
$|\chi_{1,\uparrow}\rangle, |\chi_{1,\downarrow}\rangle,
|\chi_{2,\uparrow}\rangle,|\chi_{2,\downarrow}\rangle$ , one for each
site and spin.  The only nonzero matrix elements of the one-particle
Hamiltonian 
\begin{eqnarray}
h_{\alpha,\beta}=-t 
(1-\delta_{R_\alpha,R_\beta})\delta_{\sigma_\alpha,\sigma_\beta}
\end{eqnarray}
are those with orbitals having the same
spin $\sigma_\alpha$ and $\sigma_\beta$ but different centers
$R_\alpha$ and $R_\beta$. All nonzero elements have the value $-t$,
where $t$ is positive.
The orbital energies are chosen equal to zero.

Also the interaction tensor has a simple form, namely
 \begin{equation}
 \label{eq:W_tensordimer}
   U_{\alpha\beta,\gamma\delta}=\begin{cases}
   U &\text{if $\alpha=\gamma$, $\beta=\delta$ 
             and $R_\alpha=R_\beta$ }      \\
   0 &\text{otherwise}
   \end{cases}\;.
 \end{equation}

 Thus, the Hamiltonian for the Hubbard dimer is
 \begin{eqnarray}
 \label{eq:hubbarddimerHamiltonian}
 \mathcal{H}&=&-\sum_{\sigma}t
\left(\hat{c}_{1,\sigma}^{\dagger}\hat{c}_{2,\sigma}
+\hat{c}_{2,\sigma}^{\dagger}\hat{c}_{1,\sigma}\right)+\hat{W}
\end{eqnarray}
 with the interaction
 \begin{eqnarray}
 \label{eq:hubbarddimerHamiltonianinteraction}
\hat{W}&=&\frac{1}{2}\sum_{i=1}^2\sum_{\sigma,\sigma'}U
\hat{c}_{i,\sigma}^{\dagger}\hat{c}_{i,\sigma'}^{\dagger}
\hat{c}_{i,\sigma'}^{\phantom{\dagger}}\hat{c}_{i,\sigma}.
\end{eqnarray}

\subsubsection{Total energy and density matrix}
\label{sec:total_energy_and_density_matrix}
\begin{figure}[htb]
\includegraphics[width=\linewidth,height=!]{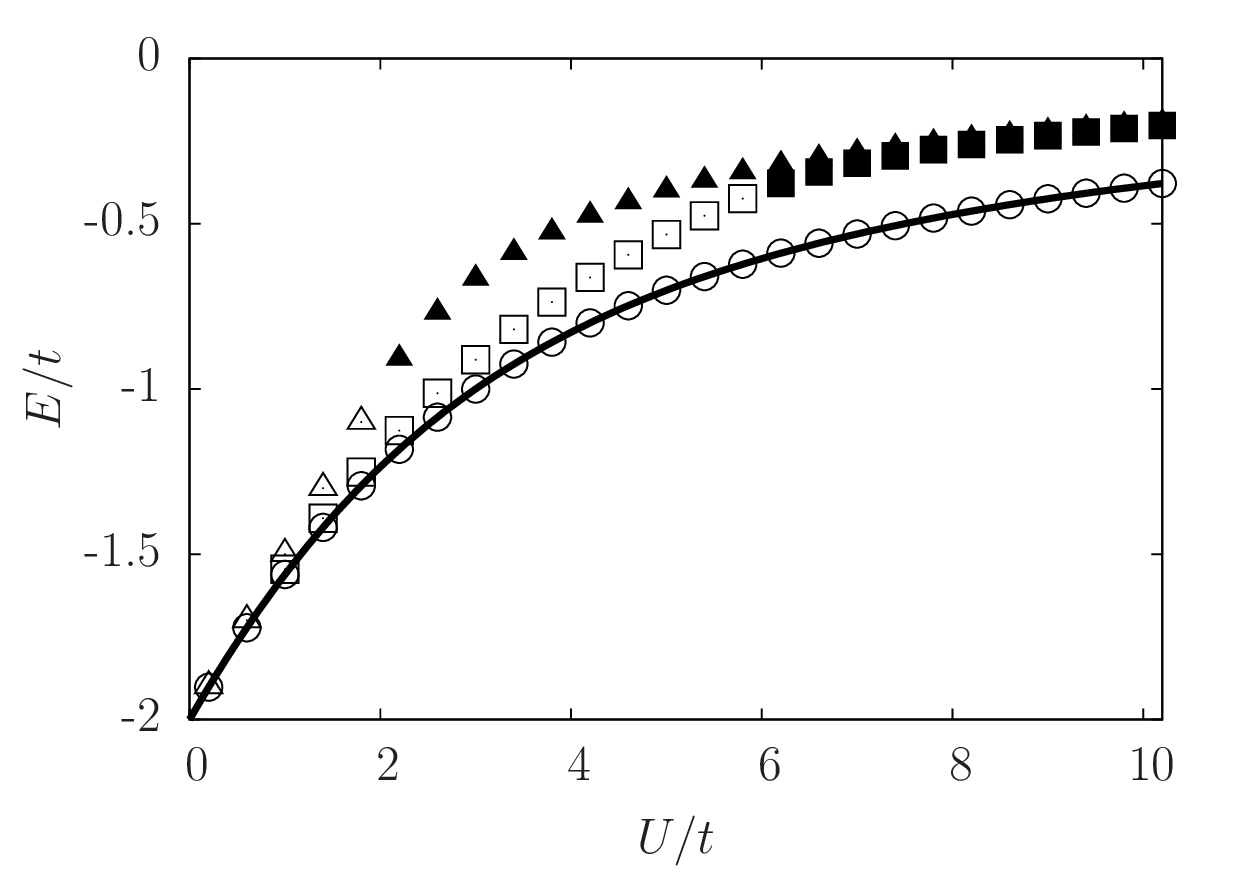}
\caption{\label{fig:Energy}Ground-state energy $E$ of the half-filled
  Hubbard dimer as function of interaction strength $U/t$ for
  different density-matrix functionals. Circles: M\"uller
  functional. Squares: power functional with $\alpha=0.53$. Triangles:
  Hartree-Fock approximation. Solid line: exact ground-state
  energy. The M\"uller functional produces the correct ground-state
  energy at half filling. Non-magnetic states are indicated by open
  symbols and antiferromagnetic states by filled symbols.}
\end{figure}

In Fig.~\ref{fig:Energy}, the total energy of the half-filled Hubbard
dimer is shown as function of interaction strength,
  alongside with the results obtained from approximate density matrix
  functionals. Some of these data have been presented
  earlier\cite{disabatino15_jcp143_24108}. Here, we emphasize the
  ground states obtained without biasing the magnetic configuration.
We follow the convention commonly adopted in the solid state community
of showing the graph for varying interaction strength $U$ and fixed
hopping $t$, so that the hopping sets the energy scale. Considering
the bond dissociation problem, the natural choice would be to keep the
interaction strength constant, while reducing the hopping parameter.

For the non-interacting case, i.e. at $U=0$, the wave function is a
Slater determinant of bonding states with opposite spin
\begin{eqnarray}
\label{eq:ground_state_limit_no_interaction}
|\Phi(U=0)\rangle=\frac{1}{2}
\left(\hat{c}^\dagger_{1,\uparrow}+\hat{c}^\dagger_{2,\uparrow}\right)
\left(\hat{c}^\dagger_{1,\downarrow}
+\hat{c}^\dagger_{2,\downarrow}\right)|\mathcal{O}\rangle\,.
\end{eqnarray}
With $|\mathcal{O}\rangle$ we denote the vacuum state. 

This wave function can be rewritten as superposition of two
eigenstates of the interaction operator
\begin{eqnarray}
|\Phi(U=0)\rangle&=&\frac{1}{2}
\left(\hat{c}^\dagger_{1,\uparrow}\hat{c}^\dagger_{1,\downarrow}
+\hat{c}^\dagger_{2,\uparrow}\hat{c}^\dagger_{2,\downarrow}\right)
|\mathcal{O}\rangle
\nonumber\\
&+&\frac{1}{2}\left(\hat{c}^\dagger_{1,\uparrow}\hat{c}^\dagger_{2,\downarrow}
-\hat{c}^\dagger_{1,\downarrow}\hat{c}^\dagger_{2,\uparrow}\right)
|\mathcal{O}\rangle.
\label{eq:psimpnonint}
\end{eqnarray}
The first wave function contains contributions with two electrons on
the same site, i.e. ionic states. Its interaction eigenvalue is
$U$. The second wave function describes two electrons with opposite
spin on different sites.  Its interaction eigenvalue is zero.

The first term describes the double occupancy, that is the probability
that two electrons are on the same site, which is penalized by the
electron-electron interaction. The second term is attributed to
left-right correlation, as it describes the probability that the two
electrons are on different sites.

As the interaction strength is increased, the contribution of the
first wave function, being responsible for double occupancy, is
suppressed. The wave function obtains the form
\begin{align}
|\Phi(\vartheta)\rangle&=&\frac{1}{\sqrt{2}}
\left(\hat{c}^\dagger_{1,\uparrow}\hat{c}^\dagger_{1,\downarrow}
+\hat{c}^\dagger_{2,\uparrow}\hat{c}^\dagger_{2,\downarrow}\right)
|\mathcal{O}\rangle
\cos\left(\vartheta+\frac{\pi}{4}\right)
\nonumber\\
&+&\frac{1}{\sqrt{2}}
\left(\hat{c}^\dagger_{1,\uparrow}\hat{c}^\dagger_{2,\downarrow}
-\hat{c}^\dagger_{1,\downarrow}\hat{c}^\dagger_{2,\uparrow}\right)
|\mathcal{O}\rangle
\sin\left(\vartheta+\frac{\pi}{4}\right)\,.
\label{eq:phiexactofalpha}
\end{align}

With a basis set in the order $
(|\chi_{1,\uparrow}\rangle,|\chi_{1,\downarrow}\rangle
,|\chi_{2,\uparrow}\rangle,|\chi_{2,\downarrow}\rangle)$,
the one-particle reduced density matrix has the form
\begin{equation}
\rho_{\alpha,\beta}(\vartheta)=\frac{1}{2}
\left(\begin{array}{cccc}
1&0&\cos(2\vartheta)&0\\
0&1&0&\cos(2\vartheta)\\   
\cos(2\vartheta)&0&1&0\\
0&\cos(2\vartheta)&0&1
\end{array}\right).
\end{equation}

The interaction energy is proportional to the double occupancy
\begin{eqnarray}
  \langle \Phi(\vartheta) |\hat W|\Phi(\vartheta)\rangle
=U\cos^2\left(\vartheta+\frac{\pi}{4}\right)
\label{eq:eintexact}
\end{eqnarray}
and the non-interacting energy is 
\begin{eqnarray}
\langle\Phi(\vartheta)|\hat{h}|\Phi(\vartheta)\rangle
=-2t\cos\left(2\vartheta\right).
\label{eq:e1pexact}
\end{eqnarray}

The value of $\vartheta$ results from an equilibrium between the
forces from the interaction energy Eq.~\eqref{eq:eintexact} and those
from the one-particle energy Eq.~\eqref{eq:e1pexact}, which determines
$\vartheta(U)$ as 
\begin{equation}
\label{eq:alphaU}
\vartheta(U)=\arctan\left(\sqrt{1+\left(\frac{U}{4t}\right)^2}
+\frac{U}{4t}
\right)-\frac{\pi}{4}\,.
\end{equation}
The value $\vartheta(U)$ varies from zero to $\pi/4$ with increasing
interaction strength.

The resulting optimum energy has the form
\begin{eqnarray}
E=-2t\left[\sqrt{1+\left(\frac{U}{4t}\right)^2}-
\frac{U}{4t}\right].
\end{eqnarray}

As the interaction increases, the wave function changes continuously
from a Slater determinant of bonding states Eq.~\eqref{eq:psimpnonint}
at $U=0$ to a singlet state with antiferromagnetic
correlations. During this process, the bond strength is weakened and
the covalent bond vanishes completely in the limit of infinite
interaction. This loss of covalent bonding can also be described as
localization of electrons on opposite sites, which raises the kinetic
energy as a consequence of Heisenberg's uncertainty principle.

What has been described here is what is called static
correlation\cite{cohen12_cr112_289}: The states for finite interaction
can no more be described by a single Slater determinant, but four
Slater determinants are required. 

\subsubsection{Natural orbitals and occupations}
Interestingly, the natural orbitals do not depend on the
interaction strength $U$. They are the bonding and antibonding states
\begin{eqnarray}
|b,\sigma\rangle
&:=&\frac{1}{\sqrt{2}}
\left(|\chi_{1,\sigma}\rangle+|\chi_{2,\sigma}\rangle\right)
\nonumber\\
|a,\sigma\rangle
&:=&
\frac{1}{\sqrt{2}}
\left(|\chi_{1,\sigma}\rangle-|\chi_{2,\sigma}\rangle\right)
\;.
\label{eq:NO}
\end{eqnarray}
Both orbitals are spread over both atoms, and the natural orbitals are
identical to those of the non-interacting system.

\begin{figure}[htb]
 \includegraphics[width=\linewidth,height=!]{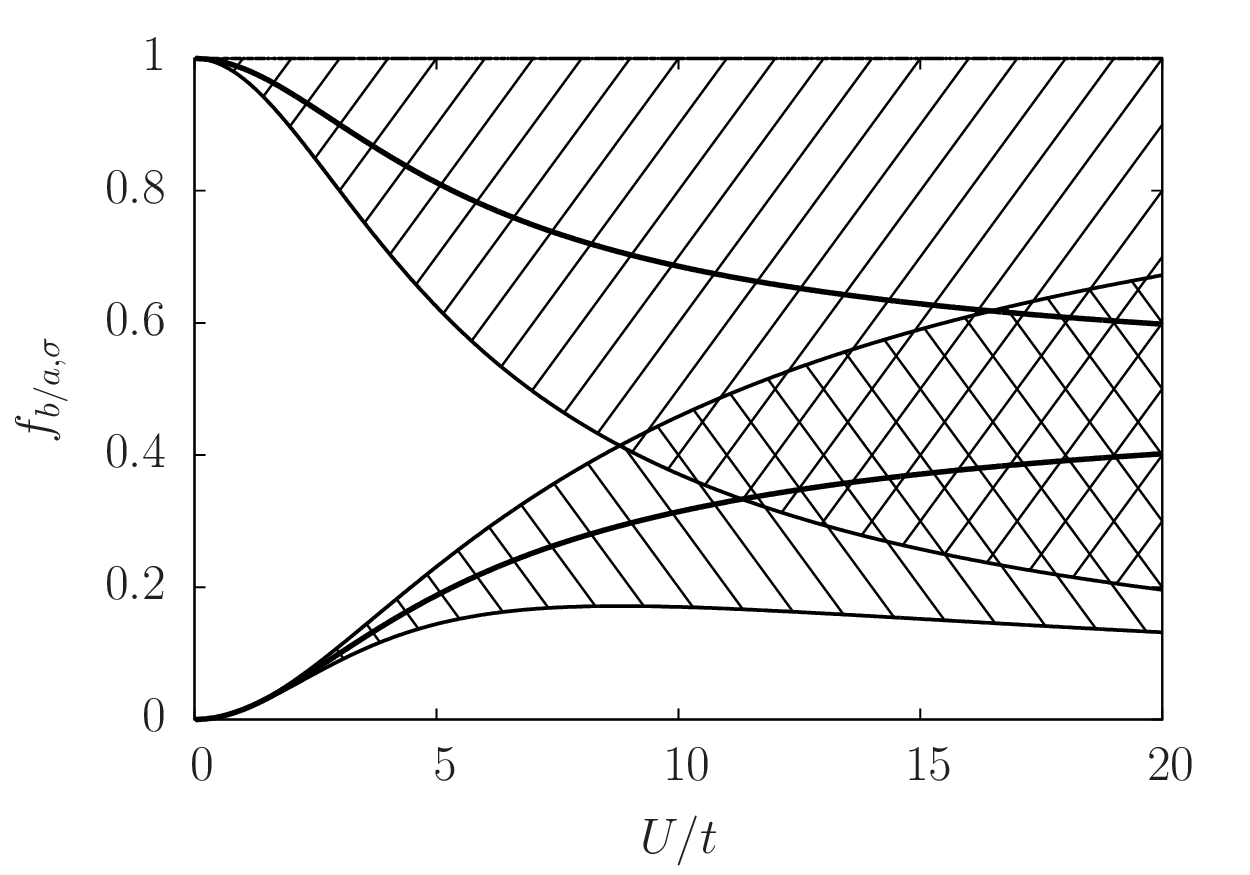}
 \caption{\label{fig:range}Occupations $f_{b,\sigma}$ and
    $f_{a,\sigma}$ from Eq.~\eqref{eq:muellersocc} of the half-filled
    dimer obtained with the M\"uller functional as a function of
    interaction strength $U/t$. The striped regions indicate the range
    of occupations in the manifold of degenerate ground-states.  The
    thick solid lines indicate the mean values for the pair of
    occupations in the corresponding striped region. It also
    represents the degenerate occupations for the non-magnetic
    solution of the M\"uller functional. The occupations of the
    non-magnetic solution of the M\"uller functional coincide with
    those of the exact ground state of the Hubbard
    dimer.}
\end{figure}

The loss of bonding is, however, expressed by the fact that the
occupations become fractional.  The occupations are shown in
Fig.~\ref{fig:range}. Their
exact values $f_{b,\sigma}$ for the bonding states and $f_{a,\sigma}$
for the antibonding states are
\begin{eqnarray}
f_{b,\sigma}&=&\frac{1}{2}+\frac{1}{2}\cos(2\vartheta)
\nonumber\\
f_{a,\sigma}&=&\frac{1}{2}-\frac{1}{2}\cos(2\vartheta).
\end{eqnarray}
In the non-interacting case, the occupations are integer, with filled
bonding states and unoccupied antibonding states. In the limit of
large interaction strength the occupations approach $\frac{1}{2}$ for
all four natural orbitals. In this limiting case with equally occupied
bonding and antibonding states, the net bond strength vanishes
completely.  In the context of natural orbitals, we describe the
effect as quantum fluctuations that create electron-hole
pairs. These electron-hole pairs destroy the covalent bond with
increasing interaction.

\subsubsection{Correlations}
\begin{figure}[htb]
 \includegraphics[width=\linewidth,height=!]{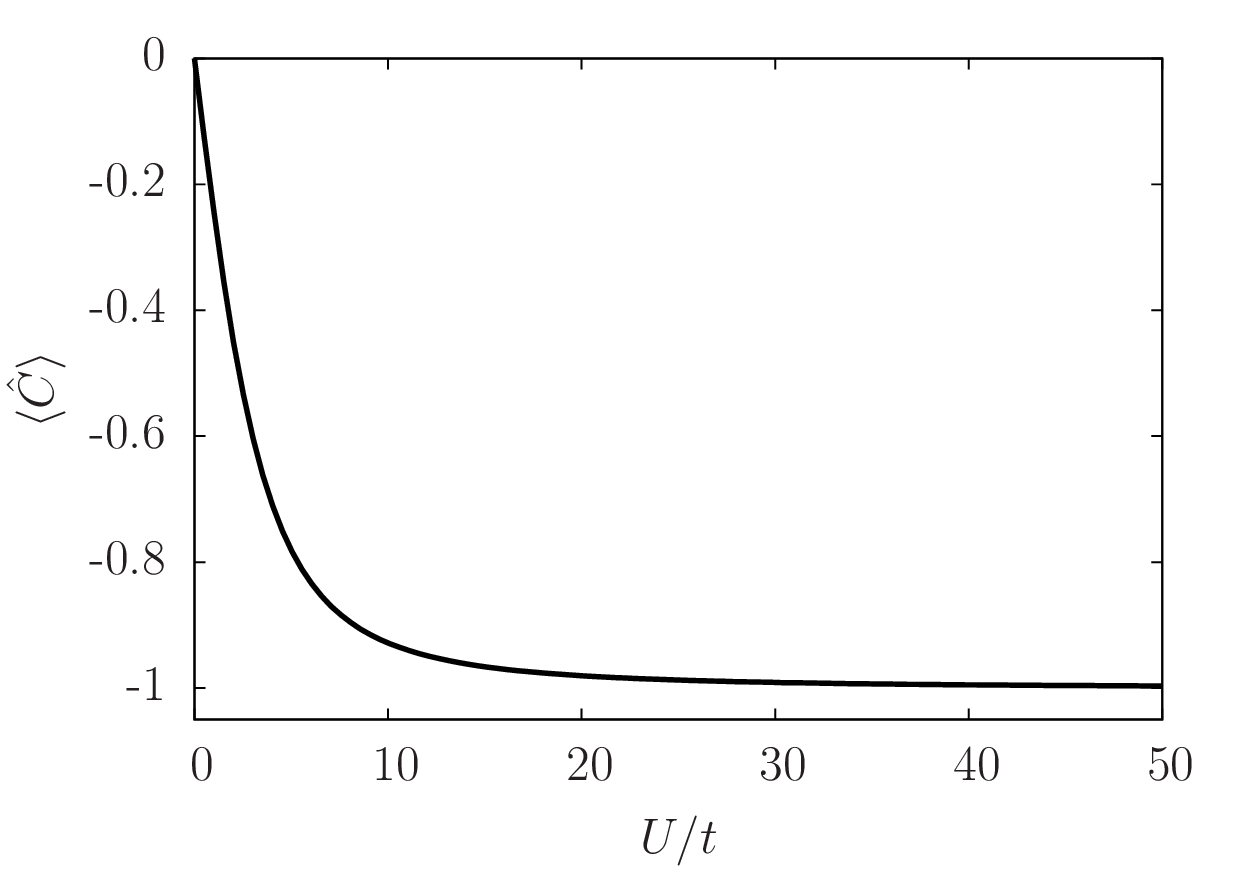}
 \caption{\label{fig:cor}Site correlation $\langle \hat{C}\rangle$ as
    defined in Eq.~\eqref{eq:sitecorrelation} of the half-filled
    Hubbard dimer as a function of $U/t$. With increasing interaction
    strength $U/t$ the site correlation shows the transition from
    delocalized electrons $\langle \hat{C}\rangle=0$ to the left-right
    correlated state with $\langle \hat{C}\rangle=-1$. }
\end{figure}

In view of the following discussion, it is instructive to investigate
the correlations of the electrons. The probability for an electron to
be on one site and the other on the other site, we name it ``site
correlation'', is given by the expectation value of
\begin{eqnarray}
\hat{C}=\left(\hat{c}^\dagger_{2,\uparrow}\hat{c}_{2,\uparrow}
-\hat{c}^\dagger_{1,\uparrow}\hat{c}_{1,\uparrow}\right)
\left(\hat{c}^\dagger_{2,\downarrow}\hat{c}_{2,\downarrow}
-\hat{c}^\dagger_{1,\downarrow}\hat{c}_{1,\downarrow}\right).
\label{eq:sitecorrelation}
\end{eqnarray}
For a state where both electrons bunch on one site, the expectation
value of this operator is one, while if they localize on opposite
sites, the expectation value is minus one. If it is zero, then the
electrons are delocalized, i.e there is no correlation between the
positions of both electrons. The correlation operator $\hat{C}$ is a
two-particle operator and is not accessible from the one-particle
density matrix. The exact solution for the correlation expectation
value for the ground state is given by
\begin{equation}
 \label{eq:corr}
 \langle \hat{C}\rangle=-\sin\Bigl(2\vartheta(U)\Bigr),
\end{equation}
where $\vartheta(U)$ is given by Eq.~\eqref{eq:alphaU}.  We can see in
Fig.~\ref{fig:cor} that the site correlation vanishes without
interaction, while the electrons anti-bunch for strong correlation so
that $\langle\hat{C}\rangle$ approaches minus one. A site correlation
of minus one indicates that each electron is fully localized either at
one or at the other site, while the other is always at the other
site. This is the basic notion of left-right correlation.

Of interest will also be the magnetic nature of the wave functions.
The operator for the spin on site $i$ is
\begin{eqnarray}
\hat{\vec{S}}_{i}&=&\frac{\hbar}{2}
\left(\begin{array}{r}
\hat{c}^\dagger_{i,\uparrow}\hat{c}_{i,\downarrow}
+\hat{c}^\dagger_{i,\downarrow}\hat{c}_{i,\uparrow}
\\
-i\hat{c}^\dagger_{i,\uparrow}\hat{c}_{i,\downarrow}
+i\hat{c}^\dagger_{i,\downarrow}\hat{c}_{i,\uparrow}
\\
\hat{c}^\dagger_{i,\uparrow}\hat{c}_{i,\uparrow}
-\hat{c}^\dagger_{i,\downarrow}\hat{c}_{i,\downarrow}
\end{array}\right).
\end{eqnarray}

For the wave function in Eq.~\eqref{eq:phiexactofalpha}, the total
spin expectation value $\langle(\hat{\vec{S}}_1+\hat{\vec{S}}_2)^2\rangle$
vanishes, and consequently the spin expectation
value $\langle\hat{\vec{S}}_{i}\rangle$ on each site vanishes as
well. Nevertheless, the spins on different sites are
antiferromagnetically correlated, that is
\begin{eqnarray}
\langle\Phi(\vartheta)|\hat{\vec{S}}_{1}\cdot\hat{\vec{S}}_{2}
|\Phi(\vartheta)\rangle 
=-\frac{3}{8}\hbar^2\left[1+\sin(2\vartheta)\right] \le 0.
\label{eq:s1s2}
\end{eqnarray}
An antiferromagnetic correlation is already present in the
non-interacting state, which expresses the non-vanishing contribution
of the left-right correlated states to the Slater determinant built
from bonding orbitals. As the interaction increases the left-right
correlation doubles, which reflects in the increase of the
antiferromagnetic correlation expressed in Eq.~\eqref{eq:s1s2}.

\section{Methods}
\label{sec:methods}
The natural orbitals and occupations have been optimized in the
Car-Parrinello spirit\cite{car85_prl55_2471} using a fictitious
Lagrangian of the form
\begin{eqnarray}
\mathcal{L}&=&
\frac{1}{2}\sum_n m_f \dot{x}^2_{n}+
\sum_n f(x_n) m_\psi \sum_\alpha|\dot{a}_{\alpha,n}|^2 
\nonumber\\
&-&
\sum_n f(x_n)\sum_{\alpha,\beta} a_{\alpha,n} h_{\beta,\alpha} a^*_{\beta,n}
-F^{\hat{W}} [\sum_{n} a_{\alpha,n} f(x_n) a^*_{\beta,n}]
\nonumber\\
&+&\sum_{n,m}\Lambda_{m,n}
\Bigl(\sum_{\alpha} a^*_{\alpha,n}a_{\alpha,m} -\delta_{n,m}\Bigr)
+\mu\left(\sum_n f(x_n)-N\right).
\nonumber\\
\end{eqnarray}
The natural orbitals are given by the complex-valued coefficients
$a_{\alpha,n}=\langle\chi_\alpha|\phi_n\rangle$ as
\begin{eqnarray}
|\phi_n\rangle=\sum_\alpha|\chi_\alpha\rangle a_{\alpha,n}
\end{eqnarray}
and the occupations $f_n=f(x_n)$ are expressed by the real-valued
dynamical variables $x_n$ with $f(x)=[1-\cos(x)]/2$. The
orthonormality of the natural orbitals is enforced with the Lagrange
multipliers $\Lambda_{m,n}$, which form a hermitian matrix, and the
particle number is constrained with the chemical potential $\mu$.

In order to avoid any bias in our results, the wave functions and
occupations are initialized as random numbers between zero and
one. Then the constraints,
i.e. orthonormality of the natural orbitals and total particle number,
are imposed. For the actual minimization, the Euler-Lagrange equations
for the occupation variables $x_n$ and the coefficients $a_{\alpha,n}$
are propagated using the Verlet algorithm under the additional action
of a friction term. The constraints are enforced with the help of
Lagrange multipliers.\cite{ryckaert77_jcompphys23_327} The friction
term leads to energy dissipation and monotonic decrease of the total
energy until the ground state or a metastable state is reached.  The
phase space has been explored by repeating the calculation, in order
to identify the global minimum and potential degenerate ground
states.

An analytical form of the natural orbitals has been extracted by
inspection of the results obtained numerically.  The resulting ansatz
for the natural orbitals has been verified by optimizing the total
energy in this subspace, and comparing the energies. While the
numerical formulation is invariant under global spin rotations,
spatial reflection, or application of a phase factor, the analytical
results are given for a particular choice.

\section{Performance of density-matrix functionals}
\label{sec:performance}
\subsection{Hartree-Fock approximation}
After having covered the main properties of the exact ground state of
the half-filled Hubbard dimer in section~\ref{sec:hubbarddimer}, we
turn now to the results obtained from approximate density-matrix
functionals.  We begin with the Hartree-Fock approximation given in
Eq.~\eqref{eq:HFK}, which has been the starting point for the
development of other density-matrix functionals investigated in this
study as discussed in
Sec.~\ref{sec:construction_of_density_matrix_functionals}.

\subsubsection{Non-magnetic solution}
If one constrains the density matrix to remain non-magnetic, the
natural orbitals do not depend on the interaction strength. The
corresponding total energy has the form
\begin{eqnarray}
E^{HF}(U)=-2t+\frac{U}{2}.
\label{eq:hfunrestricted}
\end{eqnarray}

The energy Eq.~\eqref{eq:hfunrestricted} for the Hubbard dimer with two
infinitely separated atoms, that is in the limit of $t\rightarrow0$,
results in a non-zero energy $\frac{U}{2}$, while the correct energy
vanishes, because each isolated atom has a single electron that does
not interact with itself.  This reflects the well known difficulty of
restricted, i.e. non-spin-polarized, Hartree-Fock to describe the
dissociation of chemical bonds.

While the errors caused by non-spin polarized Hartree-Fock calculations
are severe, they are not our main concern. Today's electronic
structure calculations should consider a spin-polarization whenever a
magnetization provides a lower energy. Allowing for spin polarization,
i.e. as in unrestricted Hartree-Fock or spin-density functional
theory, improves the description dramatically.  Nevertheless, the
transition from the weakly correlated to the strongly correlated
regime still differs in many ways from the correct behavior. These
differences are of interest in the following discussion.

\subsubsection{Antiferromagnetic solution}
If one allows for general variations of the density matrix, there is a
crossover at $U=2t$ from a non-magnetic solution at small interactions
to an antiferromagnetic solution at large interactions.

One set of natural orbitals that describes the transition to the
antiferromagnetic solution beyond $U=2t$ has the form 
\begin{eqnarray}
\label{eq:HFNO}
|\phi^{HF}_1(\gamma)\rangle&=&+|b,\uparrow\rangle\cos(\gamma)
+|a,\uparrow\rangle\sin(\gamma)
\nonumber\\
|\phi^{HF}_2(\gamma)\rangle&=&+|b,\downarrow\rangle\cos(\gamma)
-|a,\downarrow\rangle\sin(\gamma)
\nonumber\\
|\phi^{HF}_3(\gamma)\rangle&=&-|b,\uparrow\rangle\sin(\gamma)
+|a,\uparrow\rangle\cos(\gamma)
\nonumber\\
|\phi^{HF}_4(\gamma)\rangle&=&+|b,\downarrow\rangle\sin(\gamma)
+|a,\downarrow\rangle\cos(\gamma)\,.
\label{eq:noaf}
\end{eqnarray}
The first two natural orbitals are occupied and the remaining two are
unoccupied.

The corresponding many-particle wave function,
\begin{eqnarray}
|\Phi^{HF}(\gamma)\rangle&=&
\left[\hat{c}^\dagger_{1,\uparrow}\cos\left(\gamma-\frac{\pi}{4}\right)
+\hat{c}^\dagger_{2,\uparrow}\cos\left(\gamma+\frac{\pi}{4}\right)\right]
\nonumber\\
&\times&
\left[\hat{c}^\dagger_{1,\downarrow}\cos\left(\gamma+\frac{\pi}{4}\right)
-\hat{c}^\dagger_{2,\downarrow}\cos\left(\gamma-\frac{\pi}{4}\right)\right]
|\mathcal{O}\rangle,
\nonumber\\
\label{eq:hfmbwv}
\end{eqnarray}
is a single Slater-determinant in the basis of the natural orbitals.
For $\gamma=0$, we recover the ground state of the non-interacting
limit given in Eq.~\eqref{eq:ground_state_limit_no_interaction}.  

The many-particle wave-function Eq.~\eqref{eq:hfmbwv} has the
one-particle reduced density matrix in the basis
$(|\chi_{1,\uparrow}\rangle, |\chi_{1,\downarrow}\rangle,
|\chi_{2,\uparrow}\rangle ,|\chi_{2,\downarrow}\rangle)$
\begin{eqnarray}
\rho^{HF}(\gamma)\!&=&\!\frac{1}{2}\!
\left(\begin{array}{cccc}
1\!+\!\sin(2\gamma)\!&0&\cos(2\gamma)&0\\
0&\!1\!-\!\sin(2\gamma)\!&0&\cos(2\gamma)\\   
\cos(2\gamma)&0&\!1\!-\!\sin(2\gamma)\!&0\\
0&\cos(2\gamma)&0&\!1\!+\!\sin(2\gamma)\!
\end{array}\right).
\nonumber\\
\label{eq:hfdenmat}
\end{eqnarray}

The interaction energy is,
\begin{eqnarray}
  \langle \Phi^{HF}(\gamma) |\hat W|\Phi^{HF}(\gamma)\rangle
&=&\frac{1}{2}U\cos^2(2\gamma)
\end{eqnarray}
and the non-interacting energy is given by
\begin{eqnarray}
 \langle \Phi^{HF}(\gamma)|\hat{h}|\Phi^{HF}(\gamma)\rangle
&=&-2t\cos(2\gamma)\,.
\end{eqnarray}

Increasing the parameter $\gamma$ in the wave function from $0$,
i.e. the non-interacting limit, allows one to trade part of the covalent
bond, i.e. the kinetic energy, for a reduction of the interaction
energy.

The total energy is minimized by
\begin{equation}
 \gamma(U)=
\begin{cases}
0 & \text{for $U \leq 2t$} 
\\
\frac{1}{2}\arccos\left(\frac{2t}{U}\right) & \text{for $U>2t$.}
\end{cases}
\label{eq:hfgammofu}
\end{equation}
For $U\leq 2t$, the system remains non-magnetic and the natural
orbitals are given by bonding and antibonding orbitals as in the case
of non-magnetic dimer. But for $U>2t$, the system becomes an
antiferromagnet, whereas the exact many-particle wave function is a
singlet with antiferromagnetic correlations.
The antiferromagnetic state is a superposition of a
singlet and a triplet wave function and thus it is not an eigenstate
of $\hat{\vec{S}}^2$. We can paraphrase it as a violation of
rotational symmetry in the spin degrees of freedom, i.e. of SU(2) spin
symmetry.

\subsection{M\"uller functional}
 \label{sec:Muller}
 M\"uller's approximation to the density-matrix functional introduced
 in Sec.~\ref{sec:construction_of_density_matrix_functionals} leads to
 the exact ground-state energy for the half-filled Hubbard dimer for
 all interaction
 strengths\cite{PhysRev.101.1730,disabatino15_jcp143_24108}.  In
 contrast to the Hartree-Fock approximation, there is no unphysical
 transition to an antiferromagnetic state.

\subsubsection{Magnetic solutions}
\label{sec:Mueller_magnetic_solutions}
Even though the M\"uller functional produces exact ground-state
energies for the half-filled Hubbard dimer, we also detected a major
flaw, namely that there is a one-dimensional manifold of magnetic
states which are degenerate to the correct non-magnetic solution. The
infinite magnetic susceptibility obtained with the M\"uller functional
is in contrast to the exact behavior: At zero temperature and finite
interaction strength, the true magnetic susceptibility vanishes 
because of the finite singlet-triplet
splitting.\cite{Suezaki1972293,PhysRevB.10.3626}

Our unbiased optimizations using the M\"uller functional result in
natural orbitals equivalent to the exact ones given in
Eq.~\eqref{eq:NO}, namely the bonding and antibonding orbitals. 

With the natural orbitals of Eq.~\eqref{eq:NO}, the total energy
for the half-filled dimer obtained from the M\"uller
functional can be expressed solely by the occupations as
\begin{equation}
 E^M
=-2t+\frac{1}{2}U
+2t\left(\sum_\sigma f_{a\sigma}\right)
-\frac{1}{2}U\sum_\sigma\sqrt{f_{a\sigma}f_{b\sigma}}.
\label{eq:Energy}
\end{equation}
The first two terms, which are independent of
the occupations, are identical to the total energy
Eq.~\eqref{eq:hfunrestricted} of the spin-restricted Hartree-Fock
approximation. If only the bonding states are occupied, the remaining
terms of Eq.~\eqref{eq:Energy} vanish and the M\"uller functional gives the
same result as the Hartree-Fock approximation.

The occupations are obtained as the minimum of Eq.~\eqref{eq:Energy}
for occupations between zero and one that add up to the total particle
number of $N=2$.  For a given interaction strength, we find that the
minimum condition does not define a single point, but a line of
degenerate states parameterized by the parameter $s$
\begin{eqnarray}
\label{eq:line_solution}
 f^M_{a,\uparrow}(s)&=&\frac{1}{1+R^2}+s ,
\nonumber\\
 f^M_{a,\downarrow}(s)&=&\frac{1}{1+R^2}-s, 
\nonumber\\
 f^M_{b,\uparrow}(s)&=&R^2\left(\frac{1}{1+R^2}+s\right), \nonumber \\
 f^M_{b,\downarrow}(s)&=&R^2\left(\frac{1}{1+R^2}-s\right),
\label{eq:muellersocc}
 \end{eqnarray}
where
$R=4t/U+\sqrt{1+\left(4t/U\right)^2}$.

The requirement, that the occupations remain between zero and one,
limits the parameter $s$ to the interval
\begin{equation}
\label{eq:set}
 s\in \left[-\frac{1}{R^2(1+R^2)},\frac{1}{R^2(1+R^2)}\right].
\end{equation}
The range of the occupations, which minimize the total energy
Eq.~\eqref{eq:Energy}, is shown in Fig.~\ref{fig:range} as a function
of interaction strength $U/t$. In the limit of infinite interaction
strengths, we have $R=1$ respectively $s\in [-1/2,1/2]$ and the
possible occupations $f^M_{a/b,\sigma}(s)=1/2+\sigma s$ cover the whole
range from zero to one.

All solutions, except the physical one with equally occupied bonding
states and equally occupied antibonding states, have a magnetic
moment. Hence, the magnetic susceptibility predicted by the M\"uller
functional is infinite for all finite interaction strengths.

The magnetization of each site in the ground state of the M\"uller
functional obtained with the occupations given by
Eq.~\eqref{eq:muellersocc} has the form
\begin{eqnarray}
  \label{eq:mag}
  m^z(s)&=&\frac{1}{2}\left[
f^M_{b,\uparrow}(s)-f^M_{b,\downarrow}(s)
+f^M_{a,\uparrow}(s)-f^M_{a,\downarrow}(s)
\right]\ \mu_B\nonumber \\
  &=&\left(1+R^2\right)s \ \mu_B
\end{eqnarray}
with the Bohr magneton $\mu_B$. It can assume any value with 
$|m_z|<1/R^2\ \mu_B$.

\begin{figure}[htb]
\includegraphics[width=\linewidth,height=!]{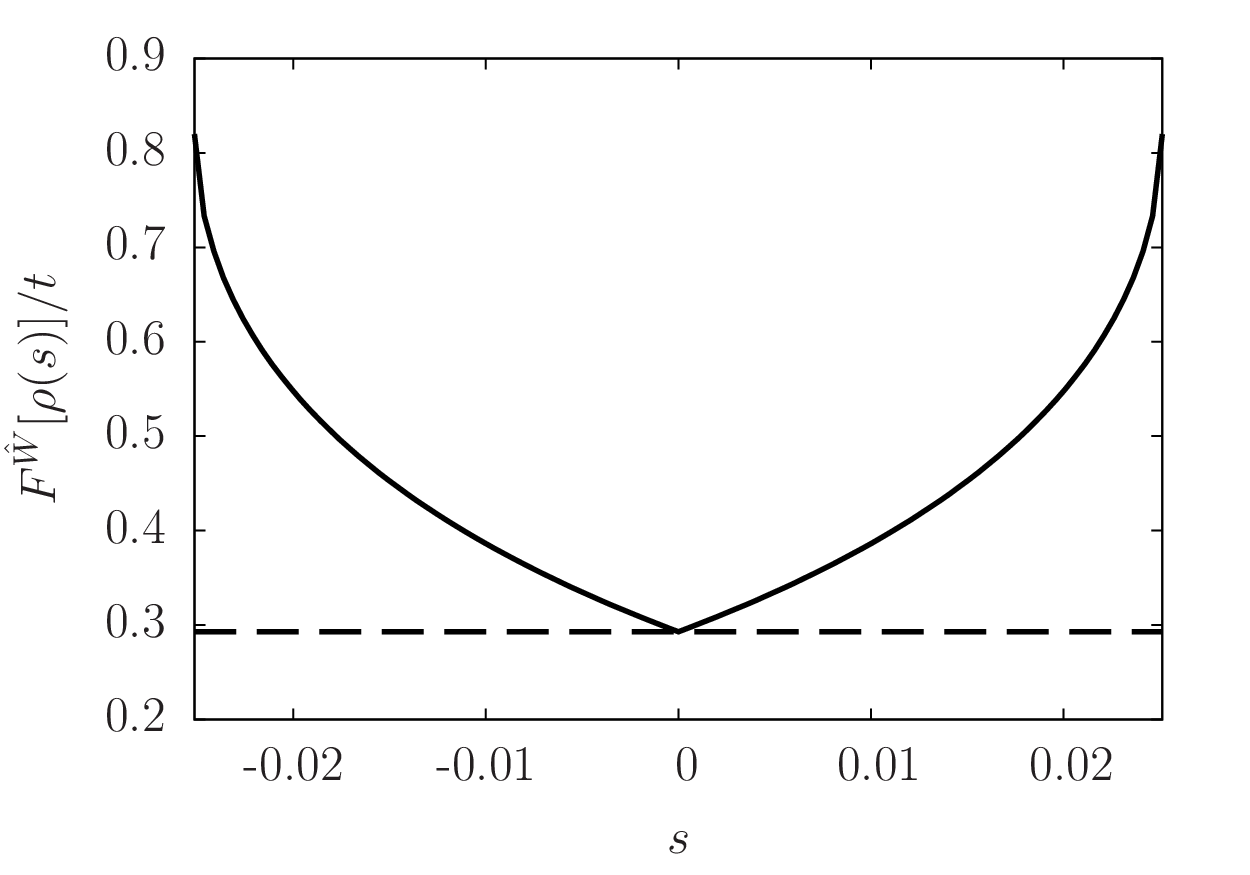}
\caption{\label{fig:comparisonMuellerExact}M\"uller density-matrix
  functional (dashed line) and the exact functional (solid line) as a
  function of the line parameter $s$ for $U=4t$. The density matrix
  $\rho(s)$ is given by Eq.~\eqref{eq:NO} and
  Eq.~\eqref{eq:line_solution}. The values of the exact functional
  have been obtained by a constrained search over an ensemble of
  many-particle wave functions. The point $s=0$, where the M\"uller
  functional and the exact functional coincide, corresponds to the
  symmetric solution ($m_z=0$).}
\end{figure}

In Fig.~\ref{fig:comparisonMuellerExact}, the density-matrix
functional of M\"uller is compared to the exact density-matrix
functional in the range of degenerate ground states of the M\"uller
functional. The exact density-matrix functional is obtained from a
constrained search over an ensemble of fermionic many-particle wave
functions\cite{bloechl11_prb84_205101} for density matrices
parametrized by Eq.~\eqref{eq:NO} and Eq.~\eqref{eq:line_solution}.
The enormous difference in the functionals illustrates the severe
problems of the M\"uller functional to describe the magnetic structure
properly and indicates a systematic flaw in the functional.

\subsubsection{Off-site interaction}
\label{sec:offsite_mueller}
The finding of an infinite magnetic susceptibility raises the question
whether this finding transfers to more realistic systems. One of the
major restrictions of the Hubbard model is the limitation to pure
on-site interactions. Therefore, we extended the Hubbard model to
include also an electron-electron interaction $V$ between the sites
\begin{eqnarray}
\label{eq:W_perturb}
\hat{W}&=&
\frac{1}{2}\sum_i
\sum_{\sigma,\sigma'}U\hat{c}^\dagger_{i,\sigma}
\hat{c}^\dagger_{i,\sigma'}\hat{c}_{i,\sigma'}\hat{c}_{i,\sigma}
\nonumber\\
&+&
\frac{1}{2}\sum_{i\neq j}\sum_{\sigma}V\hat{c}^\dagger_{i,\sigma}
\hat{c}^\dagger_{j,\sigma}\hat{c}_{i,\sigma}\hat{c}_{j,\sigma}\,.
\end{eqnarray}

Using the density matrices from the degenerate manifold of ground
states without offsite interaction, i.e. with bonding and antibonding
states as natural orbitals Eq.~\eqref{eq:NO} and the occupations from
Eq.~\eqref{eq:line_solution}, the effect of the off-site interaction
has been explored up to first order in the off-site interaction
$V$. This leads to
\begin{eqnarray}
 E^M[V,s]
 &=&E^M[0,s]
+\frac{V(R^2-1)^2}{2}
\nonumber\\
&&\times
\left[
\left(\frac{2}{1+R^2}\right)^2
+s^2
\right]+\mathcal{O}(V^2).
\label{eq:muellerenergyoffsite}
\end{eqnarray}
$E^M[0,s]$ is the $s$-independent total energy obtained with the
M\"uller functional for the Hubbard dimer in the absence of an
offsite interaction. It is given by Eq.~\eqref{eq:Energy} and Eq.~\eqref{eq:line_solution}.

\begin{figure}[htb]
\includegraphics[width=\linewidth,height=!]{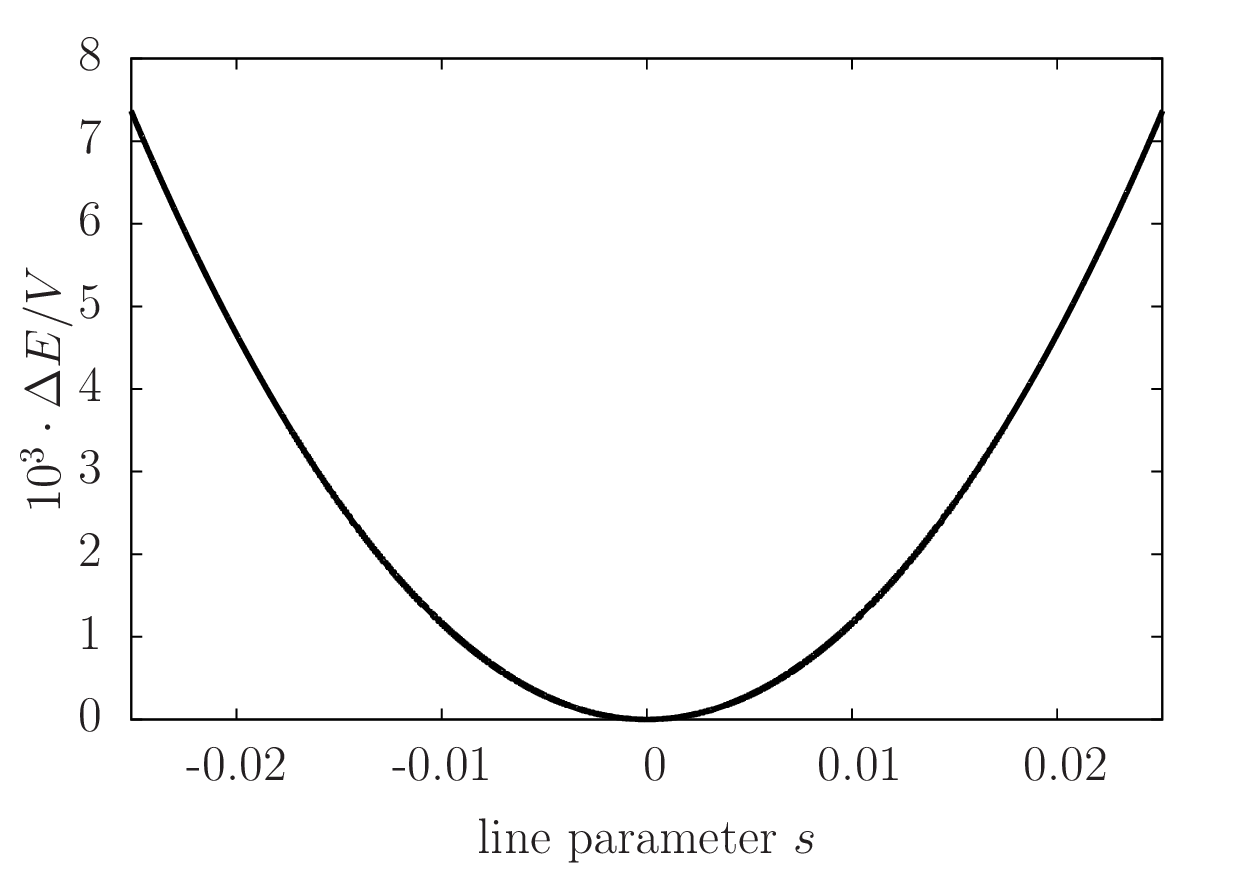}
\caption{\label{fig:Muelleroffsite}Energy $\Delta
    E=E[V,s]-E[V,s=0]$ of
  Eq.~\eqref{eq:muellerenergyoffsite} of the Hubbard dimer obtained
  with the M\"uller functional including an off-site interaction
  in first-order perturbation theory with
  $U=4t$ along the manifold Eq.~\eqref{eq:line_solution} of ground
  states. The point $s=0$ indicates the non-magnetic solution.}
\end{figure}

As shown in Fig.~\ref{fig:Muelleroffsite}, the off-site term lifts the
degeneracy of the ground states of the M\"uller functional. The
non-magnetic solution with $s=0$ is now favored. This indicates that
this artificial degeneracy may not be immediately apparent in real
systems.

Nevertheless, as evident from the comparison with the
  exact functional shown in Fig.~\eqref{fig:comparisonMuellerExact}, the
  changes produced by the off-site term are far too small: In order to produce an energy difference between the maximally
    polarized state (see Eq.~\ref{eq:set}) and the unpolarized state
    comparable to the exact result shown in
    Fig.~\ref{fig:comparisonMuellerExact}, an unrealistically large
    offsite interaction parameter of order $V=10 t$ would be
    required.

\subsection{Power functional}
After having investigated the Hartree-Fock approximation and the
M\"uller functional, we consider now the power functional invented by
Sharma et al.\cite{sharma08_prb78_201103}, which we described in
Sec.~\ref{sec:construction_of_density_matrix_functionals}.

\begin{figure}[htb]
\includegraphics[width=\linewidth,height=!]{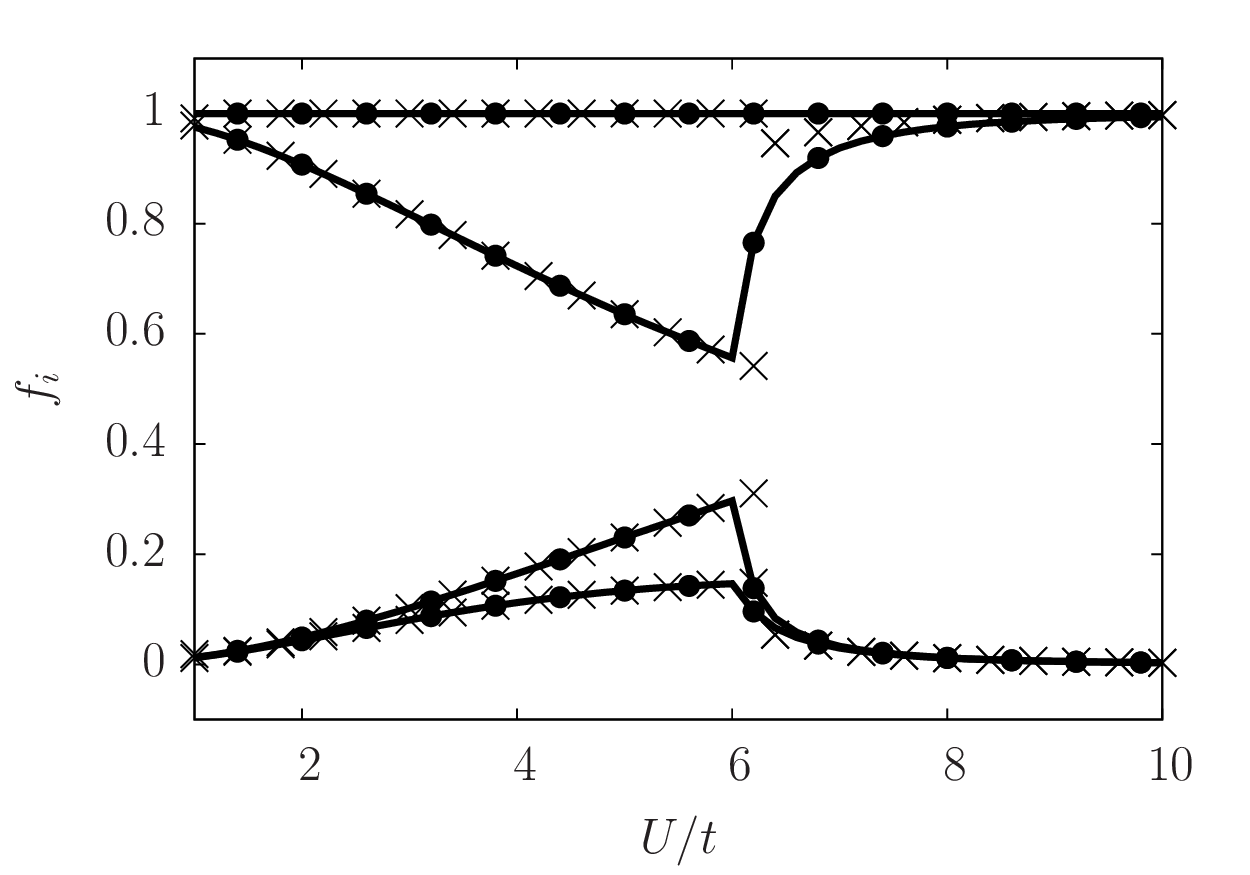}
\caption{\label{fig:power}Occupations $f_i$ as a function of $U/t$ for
  the power functional with $\alpha=0.53$. Solid dots have been
  obtained from an unbiased optimization of the power functional. The
  solid lines are obtained from a restricted optimization using the
  non-collinear natural orbitals of the ansatz
  Eq.~\eqref{eq:noncolwf}.  The diagonal crosses are obtained from a
  restricted optimization using the collinear natural orbitals
  Eq.~\eqref{eq:EAF} of the Hartree-Fock approximation.}
\end{figure}

The occupations of the Hubbard dimer obtained from the power
functional are shown in Fig.~\ref{fig:power} for a value
$\alpha=0.53$.  Whereas the density-matrix functional in the Hartree-Fock
approximation produces integer and pairwise identical occupations, the
power functional produces fractional occupations which are not
identical in pairs.

Near $U=6t$, we observe a transition. This transition separates the
M\"uller-like behavior at small interactions from a Hartree-Fock-like
behavior at large interactions.
\begin{itemize}
\item At small interactions, the solutions are analogous to those of the
  M\"uller functional. However, from the manifold of degenerate ground
  states of the M\"uller functional, the power functional favors the
  state with maximal ferromagnetic moments.
\item At larger interactions, the ground state undergoes a transition
  into a non-collinear ground state. For very large interaction the
  state approaches the Hartree-Fock-like antiferromagnetic state.
\end{itemize}

\subsubsection{Ferromagnetic solution in the weakly interacting regime:}
\begin{figure}[htb]
 \includegraphics[width=\linewidth,height=!]{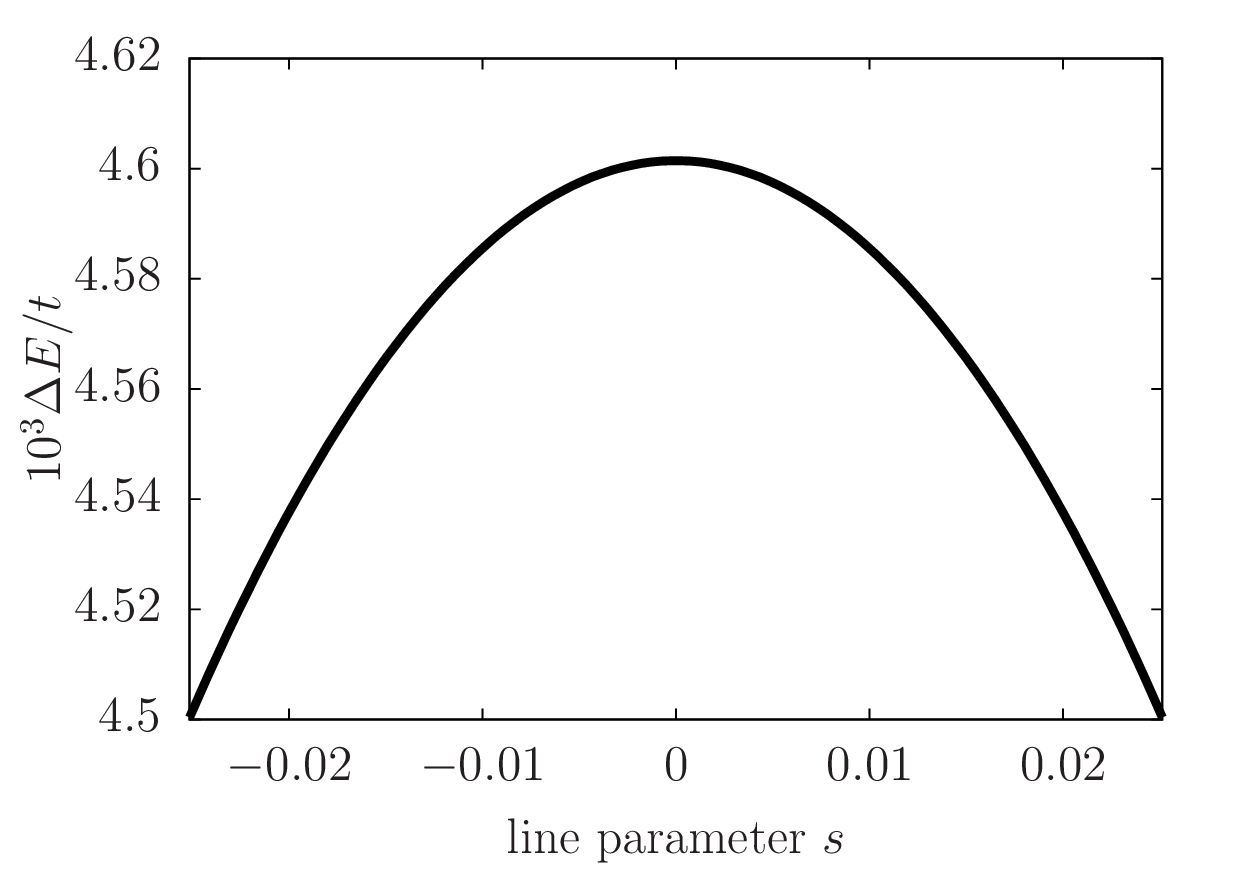}
 \caption{\label{fig:pin}Total energy difference $ \Delta
    E=E^{P}_{\alpha=1/2+\epsilon}[\rho(s)]-E^{P}_{\alpha=1/2}[\rho(s)]$
    given by Eq.~\eqref{eq:eline} for $U=4t$ using the power
    functional approximation with $\epsilon=10^{-3}$ as a function of
    the line parameter $s$ that parametrizes the one-particle reduced
    density matrix according to Eq.~\eqref{eq:line_solution}.}
\end{figure}

The occupations in the weakly interacting regime can be understood as
follows: In case of the M\"uller density-matrix functional, we have
shown in Sec.~\ref{sec:Muller} that there exists a manifold of
degenerate ground-state density matrices on the line given by
Eq.~\eqref{eq:line_solution}.  If we increase the parameter $\alpha$ of
the power functional infinitesimally as $\alpha=\frac{1}{2}+\epsilon$
where $\epsilon>0$, and restrict ourselves to interaction strengths
$U/t$ where the natural orbitals are bonding and antibonding states,
Eq.~\eqref{eq:NO}, the total energy along the line given by
Eq.~\eqref{eq:line_solution} is
\begin{eqnarray}
  \label{eq:eline} 
E^{P}_{\alpha=\frac{1}{2}+\epsilon}(s)
&=&2t\left(\frac{2}{1+R^2}-1\right)+U
\\
&-&\frac{U}{4}\sum_{\sigma=\pm 1}
\left(1+R^{1+2\epsilon}\right)^2\left(\frac{1}{1+R^2}+\sigma s\right)
^{1+2\epsilon},
\nonumber
\end{eqnarray}
where $R=4t/U+\sqrt{1+(4t/U)^2}$. The energy in Eq.~\eqref{eq:eline},
shown in Fig.~\ref{fig:pin}, has a negative curvature along the line
parameter $s$ and the minima lie at the boundaries given in
Eq.~\eqref{eq:set}.

At these boundaries, the extreme non-symmetric solutions of the
M\"uller functional, one of the states is always fully occupied (See
Fig.~\ref{fig:range}) because this maximum occupation limits the range
of degenerate solutions. This explains the corresponding observation
in Fig.~\ref{fig:power}.

Unfortunately, any change of the parameter $\alpha$ away from the
value of the M\"uller functional, destroys the non-magnetic ground
state in favor of an unphysical ferromagnetic state.

\subsubsection{Large-interaction regime}
The Hartree-Fock approximation exhibits a transition from a
non-magnetic state to an antiferromagnetic state at $U=2t$. This
transition is absent in the M\"uller functional, but it is present in
the power functional for all other values of $\alpha>\frac{1}{2}$.

In order to explore, how the power functional interpolates between
these two extreme cases, we calculated the product $\langle \hat{{\vec
    S}}_1\rangle\cdot\langle \hat{{\vec S}}_2\rangle$ of the spin
expectation values at the two sites of the dimer.  A positive value of
$\langle \hat{{\vec S}}_1\rangle\cdot\langle \hat{{\vec S}}_2\rangle$
corresponds to a ferromagnetic, a negative value to an
antiferromagnetic spin alignment. The maximum absolute value is
$\hbar^2/4$.

\begin{figure}[htb]
 \includegraphics[width=0.9\linewidth,height=!]{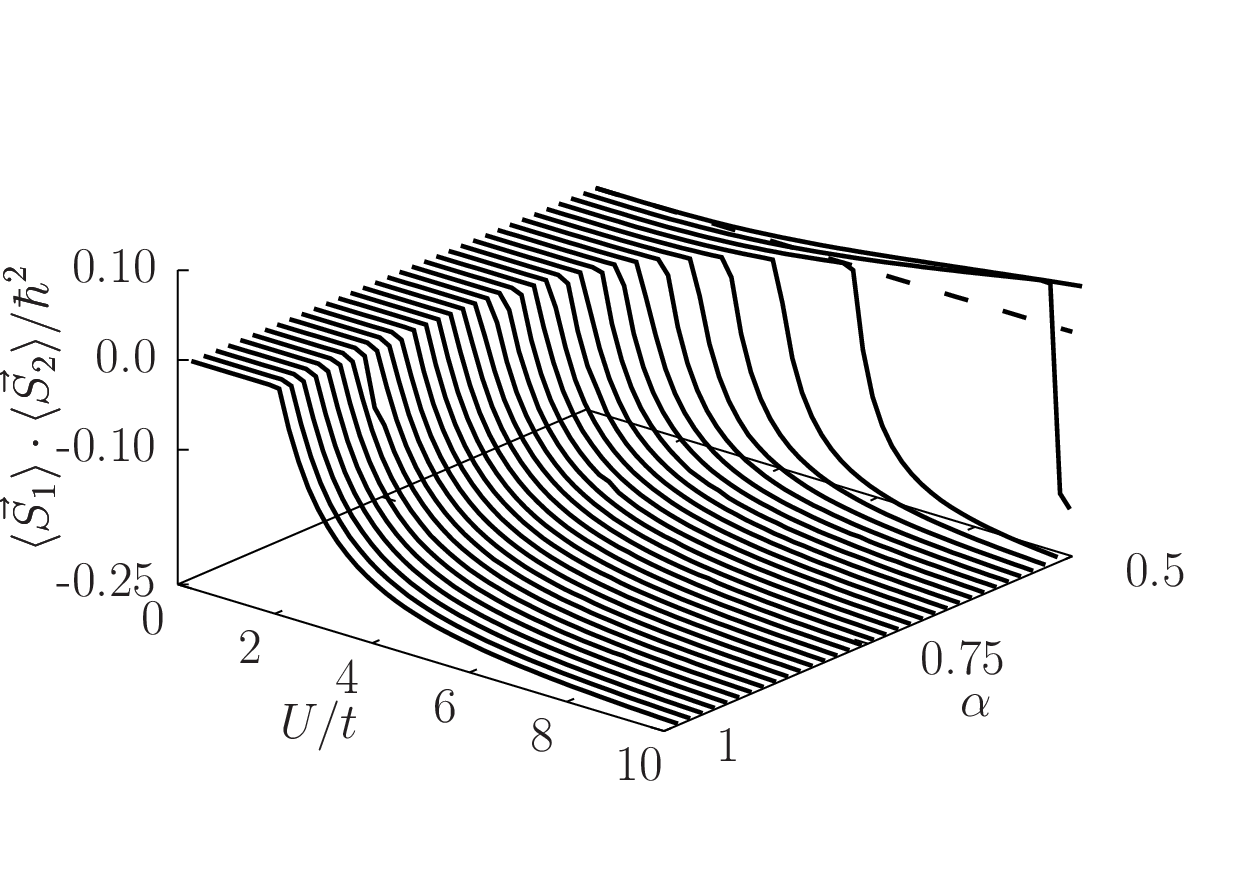}
 \caption{\label{fig:Mag}Scalar product
    $\langle\hat{\vec{S}}_1\rangle\cdot\langle\hat{\vec{S}}_2\rangle$
    of the spin expectation vectors on the two sites of the Hubbard
    dimer as an indicator for the transition to the antiferromagnetic
    state within the power functional approximation with the parameter
    $\alpha$ for the Hubbard dimer at various interaction strengths. A
    positive value indicates a ferromagnetic state, a negative value
    an antiferromagnetic state. For the M\"uller functional,
    i.e. $\alpha=1/2$, the dashed line represents the result for the
    symmetric solutions and the solid line the corresponding
    degenerate result for the degenerate maximally polarized state.  }
\end{figure}

For the Hubbard dimer at half filling, $\langle \hat{{\vec
    S}}_1\rangle\cdot\langle \hat{{\vec S}}_2\rangle$ is shown in
Fig.~\ref{fig:Mag} as function of interaction strength $U/t$ and the
parameter $\alpha$ of the power functional.  For the M\"uller
functional discussed in Sec.~\ref{sec:Mueller_magnetic_solutions}, ,
i.e. for $\alpha=1/2$, we consider the solution with the strongest
polarization, because this is the state that continuously matches to
the solutions of the power functional. In this ferromagnetic state,
$\langle \hat{\vec S}_1\rangle \cdot \langle \hat{\vec S}_2\rangle$ is
positive. Unfortunately, the correct non-magnetic state is not a
ground state of the power functional for $\alpha>\frac{1}{2}$.

\begin{figure}[htb]
\includegraphics[width=\linewidth,height=!]{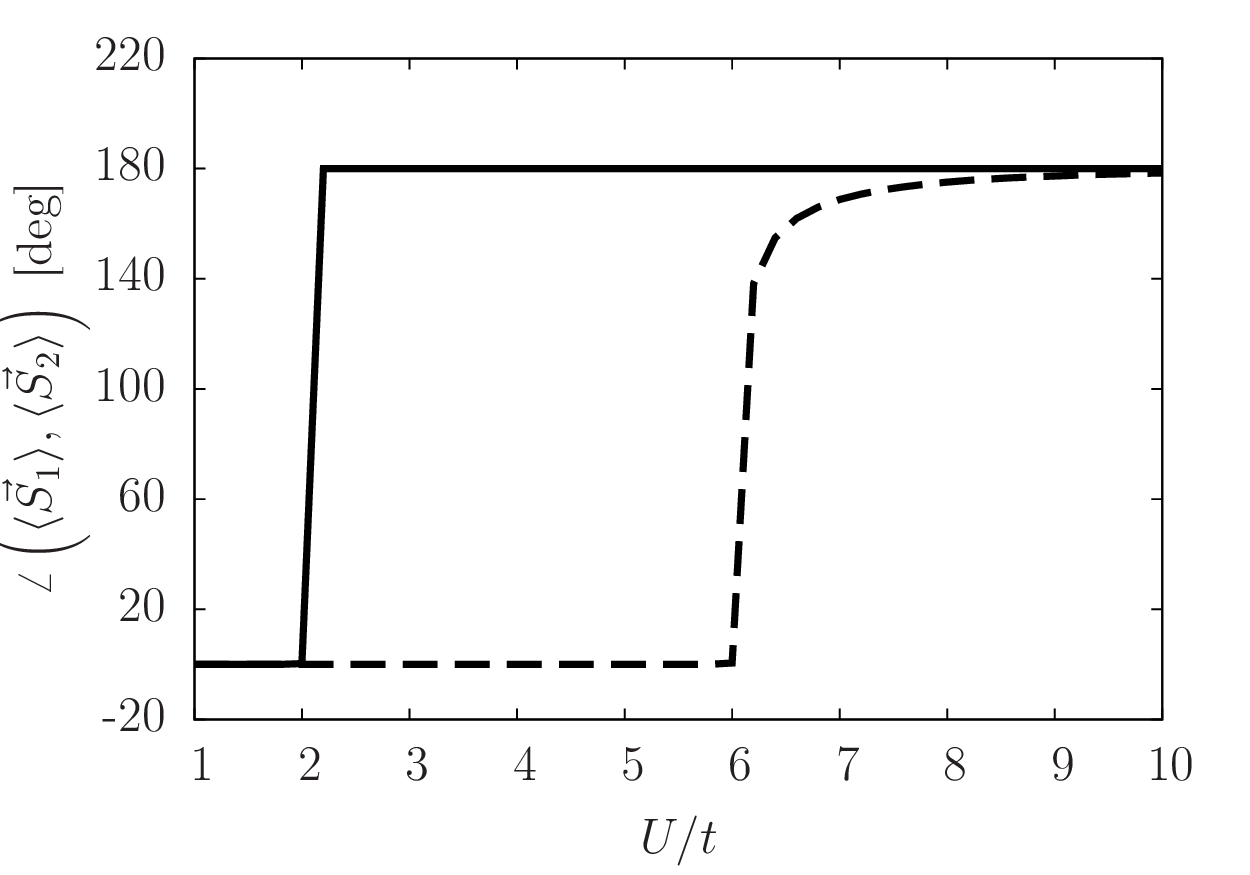}
\caption{\label{fig:noncoll}Angle between the spin expectation vectors
  $\langle\vec{S}_1\rangle$ and $\langle\vec{S}_2\rangle$ on the two
  sites of the Hubbard dimer as function of the interaction strength
  $U$.  Dashed line: power functional with the parameter
  $\alpha=0.53$; solid line: Hartree-Fock approximation.  This angle
  is a measure of collinearity of natural orbitals. }
\end{figure}

At a critical interaction strength $U_c(\alpha)$ the power functional
exhibits a transition from this ferromagnetic state into a complex
non-collinear state with a mostly antiferromagnetic spin alignment.
The angle between the magnetization on the two sites is shown in
Fig.~\ref{fig:noncoll}.

Fig.~\ref{fig:Mag} clearly shows the location of the transition
between the ferromagnetic and the antiferromagnetic non-collinear
regime.  The critical interaction strength $U_c(\alpha)$ of this
transition is infinite for the M\"uller functional. As the parameter
$\alpha$ is increased, the critical interaction strength falls off
rapidly and approaches the value $U_c(\alpha=1)=2t$ of the
Hartree-Fock approximation.

Thus, the power functional exhibits a Hartree-Fock-like transition into
an antiferromagnetic ground state except for the limiting case, the
M\"uller functional. By choosing the parameter $\alpha$ sufficiently
close to $1/2$, the transition can be shifted into a regime that is
physically less important.

\paragraph{Collinear approximation using the Hartree-Fock natural orbitals}
In order to get a qualitative understanding of the asymmetric
occupations (Ref. Fig.~\ref{fig:power}) and the critical value of
interaction strength $U_c$ of the transition to antiferromagnetic
solutions (Ref. Fig.~\ref{fig:Mag}), we use an ansatz that covers both
extreme cases, namely the M\"uller functional with $\alpha=\frac{1}{2}$
and the Hartree-Fock approximation with $\alpha=1$.  These are, one
the one hand, the asymmetric natural orbitals Eq.~\eqref{eq:HFNO} that
can describe the antiferromagnetic state of the Hartree-Fock
approximation.  On the other hand, the ansatz allows for fractional
occupations to capture the nature of the ground state of the M\"uller
functional.

With this ansatz, the one-particle reduced density matrix
$\rho(f_1,\ldots,f_4,\gamma)$ is a function of occupations $f_n$ and
the angle $\gamma$ and the corresponding total energy $E^{P,\alpha}$
obtained with the power functional is
\begin{eqnarray}
  E^{P}_{\alpha}[\rho(f_1,\ldots,f_4,\gamma)]
&=&E^{\text{kin}}[\rho(f_1,\ldots,f_4,\gamma)]
\nonumber\\
&+&F^{P}_{\alpha}[\rho(f_1,\ldots,f_4,\gamma)]
\end{eqnarray}
where
\begin{widetext}
\begin{eqnarray}
E^{\text{kin}}[\rho(f_1,\ldots,f_4,\gamma)]
&=&-t\cos(2\gamma)\left(f_1+f_2-f_3-f_4\right) \nonumber \\
F^{P}_{\alpha}[\rho(f_1,\ldots,f_4,\gamma)]
&=&\frac{U}{4}\Bigl[
(f_1+f_2+f_3+f_4)^2-(f^\alpha_1+f^\alpha_3)^2
-(f^\alpha_2+f^\alpha_4)^2\Bigr]
\nonumber\\
&+&
\frac{U}{4}\sin^2(2\gamma)
\Bigl[(f_2+f_3-f_4-f_1)^2-(f_1^\alpha-f_3^\alpha)^2
-(f^\alpha_2-f^\alpha_4)^2\Bigr].
 \label{eq:EAF}
\end{eqnarray}
\end{widetext}

An approximation, which is a strict upper bound, for the total energy
of the power functional is obtained by minimizing Eq.~\eqref{eq:EAF}
for a half-filled system with occupations between zero and one.

As a characteristic example, the resulting occupations for
$\alpha=0.53$ are shown in Fig.~\ref{fig:power}.  The properties of
this ansatz with regard to the description of the transition to the
antiferromagnetic state will be investigated in the following section
after a more general discussion of the transition.

\begin{figure}[htb]
  \includegraphics[width=\linewidth,height=!]{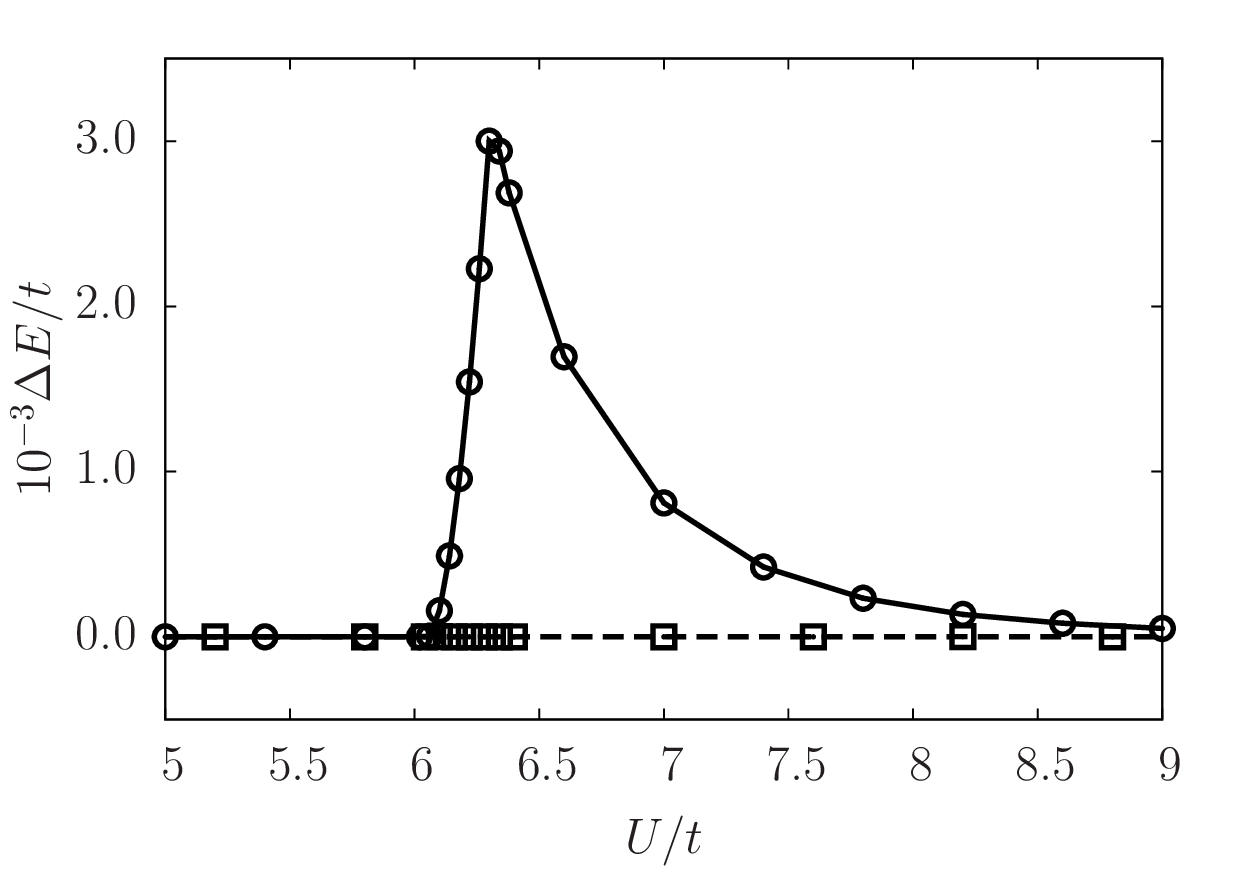}
 \caption{\label{fig:gamma}Energy difference $\Delta E$ of the power
    functional with $\alpha=0.53$ between the density matrices
    obtained by a constrained and an unbiased optimization. $\Delta E$
    for the case of constrained optimization with the natural orbitals
    of a Hartree-Fock in Eq.~\eqref{eq:HFNO} is the solid line with
    circle symbols, while the $\Delta E$ obtained from constrained
    optimization with the non-collinear natural orbitals of
    Eq.~\eqref{eq:noncolwf} is the dashed line with square symbols.
   }
\end{figure}

The ansatz using the collinear natural orbitals Eq.~\eqref{eq:HFNO} and
arbitrary occupations is, however, not able to describe the true
ground state for the power functional in the strongly interacting
regime. The energy difference of the ansatz to the unbiased solution
is shown in Fig.~\ref{fig:gamma}.  The deviation is largest near the
transition. The transition point is slightly displaced by the
collinear ansatz, which explains the sharp rise. For larger
interactions, the error due to the collinear approximation falls off
rapidly. It should be noted that the overall error due to the
restricted ansatz is apparently small. For the parameter $\alpha=0.53$
used in Eq.~\eqref{fig:gamma}, the maximum error in the energy is less
than 1~\% of the binding energy.

\paragraph{Beyond the collinear approximation}
The ansatz using the Hartree-Fock natural orbitals already gives a
fairly good description of the ground state of the power functional.
How do the natural orbitals of the power functional differ from those
of the Hartree-Fock solution?

\begin{figure}[htb]
\includegraphics[width=\linewidth,height=!]{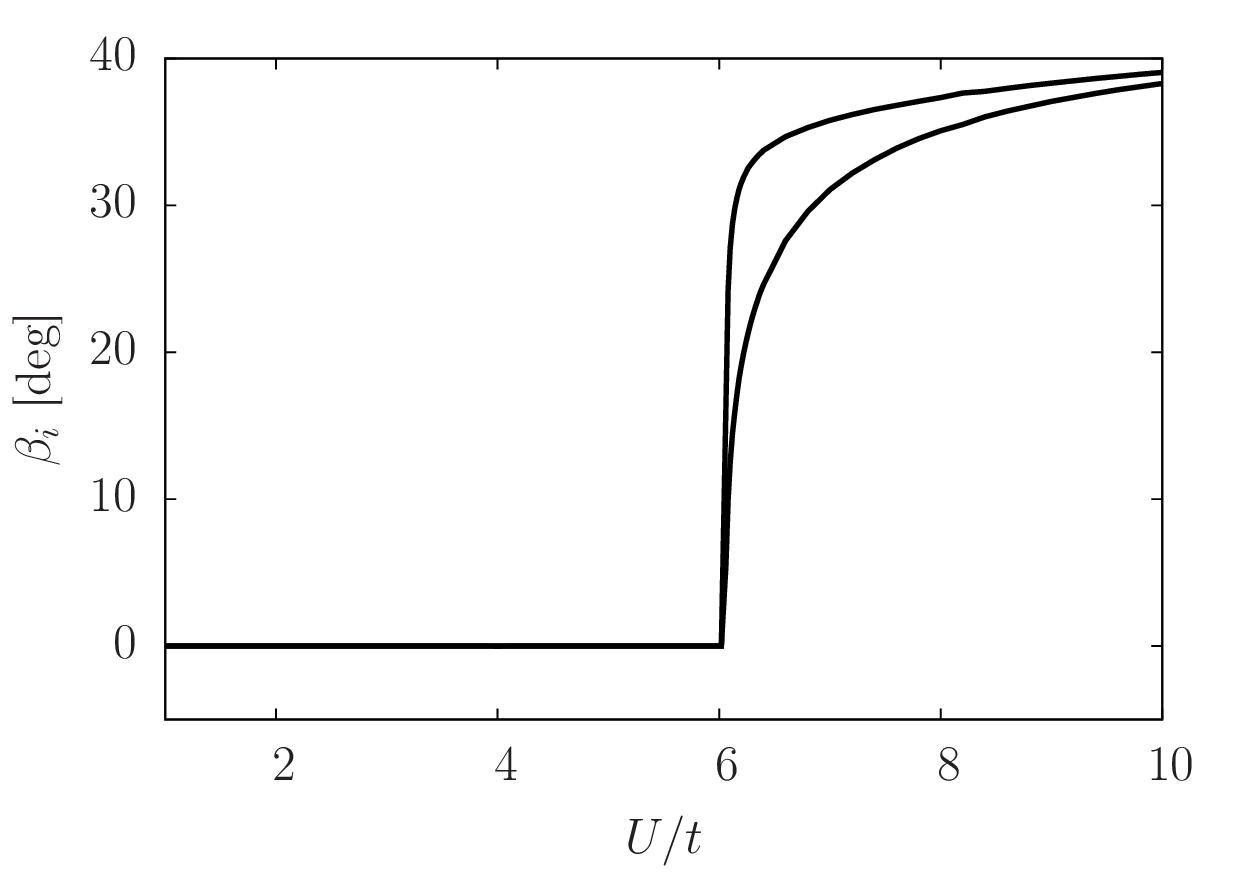}
 \caption{\label{fig:phi}Parameters $\beta_1$ and $\beta_2$ defining
    the natural orbitals Eq.~\eqref{eq:noncolwf} of the power functional
    for $\alpha=0.53$ as function of the interaction
    strength.}
\end{figure}

In the large interaction region, the power functional produces
non-collinear ground states.  The natural orbitals of the power
functional can be represented as superpositions of bonding and
antibonding states,
\begin{eqnarray}
\label{eq:noncolwf}
|\phi^P_1\rangle&=&
|b,\uparrow\rangle\cos(\beta_1)-|a,\downarrow\rangle\sin(\beta_1)
\nonumber\\
|\phi^P_2\rangle&=&|b,\downarrow\rangle\cos(\beta_2)
-|a,\uparrow\rangle\sin(\beta_2)
\nonumber\\
|\phi^P_3\rangle&=&|b,\uparrow\rangle\sin(\beta_1)
+|a,\downarrow\rangle\cos(\beta_1)
\nonumber\\
|\phi^P_4\rangle&=&|b,\downarrow\rangle\sin(\beta_2)
+|a,\uparrow\rangle\cos(\beta_2)\,.
\end{eqnarray}
The two angles $\beta_1$ and $\beta_2$ are free variational
parameters.  The natural orbitals of the non-interacting system,
respectively those of the M\"uller functional are obtained with
$\beta_1=\beta_2=0$. The values of the two parameters are shown in
Fig.~\ref{fig:phi} for one example of the power functional.

In the Hartree-Fock approximation, respectively in the power functional
with the collinear ansatz, the pair of bonding and antibonding
orbitals that contribute to a natural orbital, given in
Eq.~\eqref{eq:HFNO}, have the same spin direction. This results in the
localization of the electron on one or the other site of the
dimer. The emerging picture is appealing because it reflects the
left-right correlations of the electrons.  The admixture of
antibonding states to the natural orbitals for the two spin directions
is the same. Thus, there is no symmetry-breaking charge
disproportionation.

The natural orbitals Eq.~\eqref{eq:noncolwf} of the power functional
are composed of bonding and antibonding orbitals with opposite spin
directions. This leads to natural orbitals with equal weight on both
sites, but the spins on both sides have a finite angle between
them. The state has an intrinsically non-collinear, even though still
a coplanar spin structure.

The admixture of antibonding states in the two pairs is independent in
the power functional, so that the natural orbitals contain two
independent free parameters, namely $\beta_1$ and $\beta_2$.

The net magnetic moment of each of the four natural orbitals points
along the same direction. For the choice in Eq.~\eqref{eq:noncolwf},
this is the $z$-direction. The parameters $\beta_1$ and $\beta_2$
control the relative angles of the local moments on the two sites of
the dimer for each of the natural orbitals. This angle is $4\beta_1$
for the orbitals $|\phi^P_1\rangle$ and $|\phi^P_3\rangle$ and it is
$4\beta_2$ for the orbitals $|\phi^P_2\rangle$ and $|\phi^P_4\rangle$.
The natural orbitals are pairwise antiparallel: On any given site
$|\phi^P_1\rangle$ and $|\phi^P_3\rangle$ have local moments in
opposite directions. Similarly, this holds for $|\phi^P_2\rangle$ and
$|\phi^P_4\rangle$.

It seems that the ground states of the power functional do not connect
continuously to those of the Hartree-Fock approximation, because the
natural orbitals belong to different classes. This is, however, not
so: The ansatz for the natural orbital Eq.~\eqref{eq:noncolwf} connects
smoothly to those of the Hartree-Fock approximation in
Eq.~\eqref{eq:HFNO} when the two parameters $\beta_1$ and $\beta_2$
become equal, and furthermore the occupations become integer. This
limit of the ansatz Eq.~\eqref{eq:noncolwf} for the power functional
describes, however, an antiferromagnet with the local moments aligned
along the $x$-direction, while the ansatz of Eq.~\eqref{eq:HFNO} for the
Hartree-Fock solution is polarized along the z-direction. Thus they
are related by a global spin rotation, which is a symmetry of the
Hamiltonian.

\section{Beyond half-filling}
 \label{sec:beyondhalffilling}
Up to now, we considered only the half filled case of the Hubbard dimer.
Here we consider also deviations from the particle number $N=2$.

To avoid mathematical complications, we define $E(N)$
thermodynamically consistent as the zero-temperature limit of the
Helmholtz potential $\beta\rightarrow\infty$, which in turn is
constructed from the grand potential by a Legendre-Fenchel transform
\begin{equation}
E(N)=\lim_{\beta\rightarrow\infty}\max_\mu \left[-\frac{1}{\beta}
\ln\left(\Tr\;\e{-\beta(\hat{H}-\mu\hat{N})}\right)+\mu N\right].
\end{equation}
The trace is performed over the fermionic Fock space. 

It can be shown that the total energy $E(N)$ consists of piecewise
linear segments between integer particle numbers. Thus the slope of
the total energy $E(N)$, the chemical potential $\mu=dE/dN$, is
usually\footnote{For degenerate states and when the electron addition
  and removal are dominated by delocalized states the discontinuity
  may also vanish or become infinitesimally small.}  discontinuous at
integer occupations.  This derivative discontinuity gives the
fundamental band gap which is defined as the difference of electron
affinity and ionization potential. The band gap is relevant, not only
as an estimation related to optical spectra, but, more importantly,
for the response functions and chemical equilibria. Therefore, we
investigate whether the derivative discontinuities are properly
described by the approximate density-matrix functionals.

\begin{figure}[htb]
  \includegraphics[width=\linewidth,height=!]{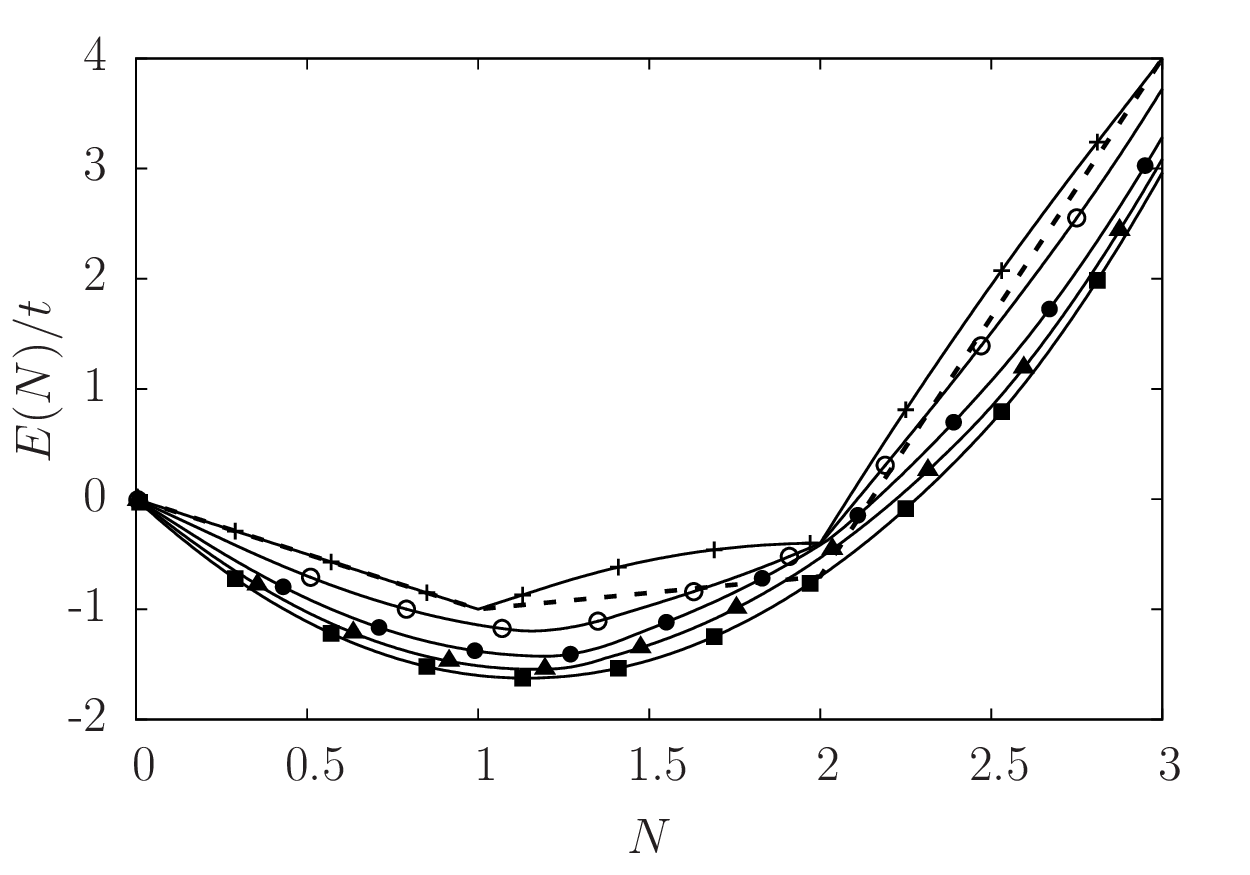}
  \caption{\label{fig:DerDiscon} Ground-state energy $E(N)$ of the
    Hubbard dimer with $U=5t$ in units of the hopping parameter $t$ as
    function of particle number $N$. The critical power functional parameter for the transition to an antiferromagnetic state lies at $\alpha\approx0.54$ for the given interaction strength. Dashed line: exact solution,
    crosses: Hartree-Fock approximation, open circles:
    power functional with $\alpha=0.7$, filled circles: power functional with $\alpha=0.58$, triangles: power functional with $\alpha=0.53$, squares: M\"uller functional.   }
\end{figure}

\begin{figure}[htb]
  \includegraphics[width=\linewidth,height=!]{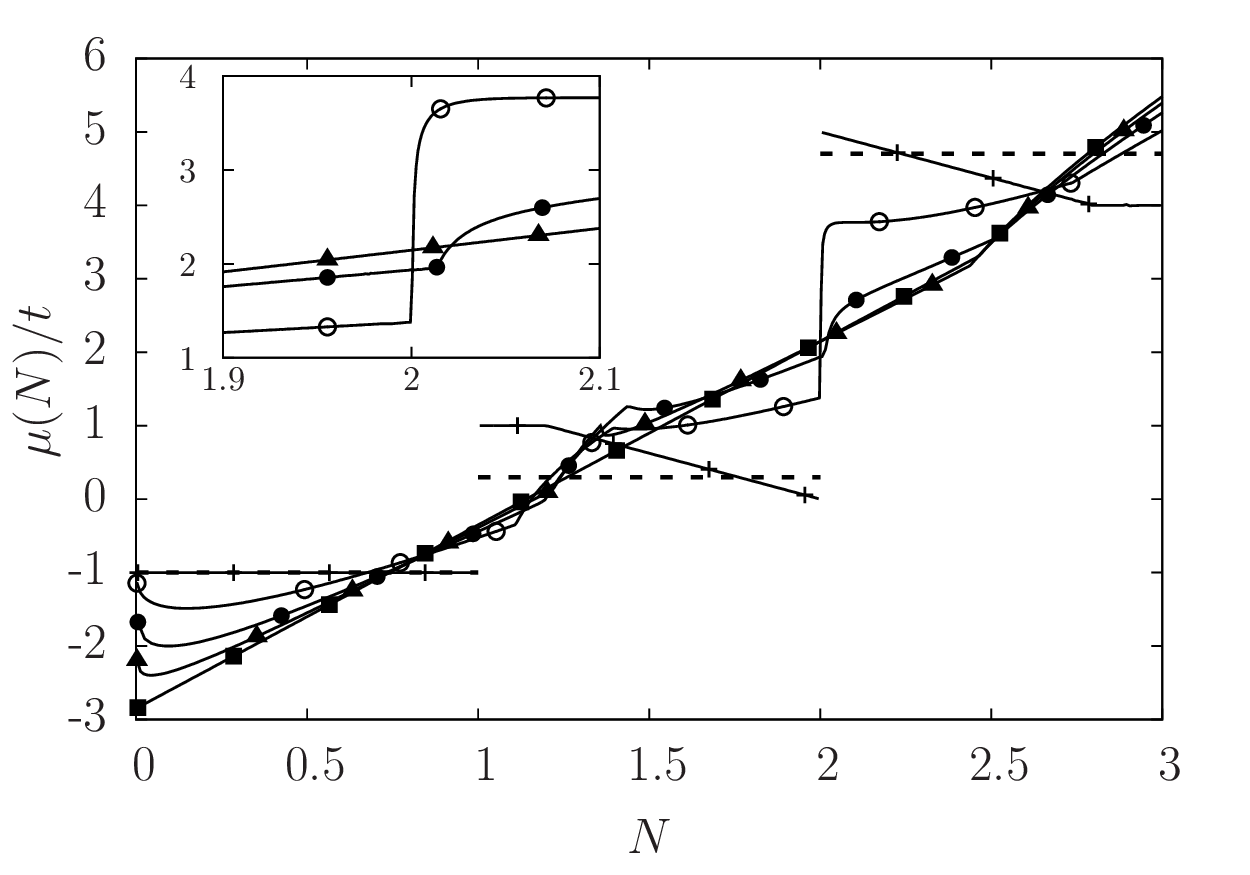}
  \caption{\label{fig:DerDiscon_mu}Chemical potential $\mu(N)$ of
      the Hubbard dimer with $U=5t$ in units of the hopping parameter
      $t$ as function of particle number $N$. The behavior of the
      power functional with $1/2<\alpha<1$ close to half-filling is
      shown in the inset.  Dashed line: exact solution, crosses:
      Hartree-Fock approximation, open circles: power functional with
      $\alpha=0.7$, filled circles: power functional with
      $\alpha=0.58$, triangles: power functional with $\alpha=0.53$,
      squares: M\"uller functional.  }
\end{figure}

The total energy $E(N)$ of the exact solution and several power functionals is shown in Fig.~\ref{fig:DerDiscon} and the corresponding chemical potential in Fig.~\ref{fig:DerDiscon_mu}. For the
Hubbard dimer, the derivative discontinuity at $N=2$ is due to a
combination of the one-particle gap and the interaction.  The
derivative discontinuity at $N=1$ is, however, entirely due to the
interaction.  These features are clearly visible for the exact
calculation shown in Fig.~\ref{fig:DerDiscon}.

In the Hartree-Fock approximation, the energy for fractional
occupations has a negative curvature for $1<N<3$. As a result, the
derivative discontinuities are larger than in the exact
solution. It reflects the well known
observation that Hartree-Fock calculations overestimate band
gaps. This observation can be rationalized with a lack of screening in
the Hartree-Fock approximation that reduces the effective interaction
strength.

The M\"uller functional, however, fails to give any derivative discontinuity.
It is surprising,
that a functional that performs as well as the M\"uller functional for
$N=2$ is completely unable to capture the correct physics beyond half
filling. It adds to the simplified picture that the M\"uller
functional behaves very metal-like: It does not have a band gap and
and its magnetic susceptibility is infinite. 

Except for the Hartree-Fock limit, also the power
  functional lacks a derivative discontinuity.  This is apparent from
  Fig.~\ref{fig:DerDiscon_mu}. For small $\alpha$, that is the
  M\"uller-like regime, the power functional behaves analogous to the
  M\"uller functional itself.  In the parameter regime of the
  antiferromagnetic ground state, however, the chemical potential
  makes a continuous transition between two distinct linear functions
  of the particle number.

This behavior of the power functional for the Hubbard dimer
  is analogous to that observed earlier for
  finite\cite{PhysRevA.79.022504,
    Z._Phys._Chem_224_467} and extended
  systems\cite{sharma08_prb78_201103,EPL_77_(2007)_67003}.

  In order to extract values for the band gap despite of the
    absence of a derivative discontinuity, Sharma et
    al.\cite{sharma08_prb78_201103} proposed the extrapolation method,
    which exploits the behavior of $E(N)$ further away from the Fermi
    level.  Sharma et al. exploit that the chemical potential makes a
    transition between two linear functions. The extrapolation of
    these linear functions to the integer particle number yields an
    offset which is identified with the band gap. This method yields
    finite band gaps in the appropriate parameter range of the power
    functional, where the M\"uller functional incorrectly predicts a
    vanishing band gap\cite{sharma08_prb78_201103}. Surprisingly, the
    band gaps obtained using the extrapolation method from the power
    functional agree well with experimental results even for
    non-magnetic calculations.

    Our results for the Hubbard dimer shown in
    Fig.~\ref{fig:DerDiscon_mu} indicate that the band gap obtained
    with the extrapolation method\cite{sharma08_prb78_201103} can be
    tuned between zero and the Hartree-Fock value by adjusting
    $\alpha$. Signatures of this behavior have been observed in
    studies that investigated the dependence on the power functional
    parameter $\alpha$ for realistic
    systems\cite{sharma08_prb78_201103,Z._Phys._Chem_224_467}.
  
\begin{figure}[htb]
  \includegraphics[width=\linewidth,height=!]{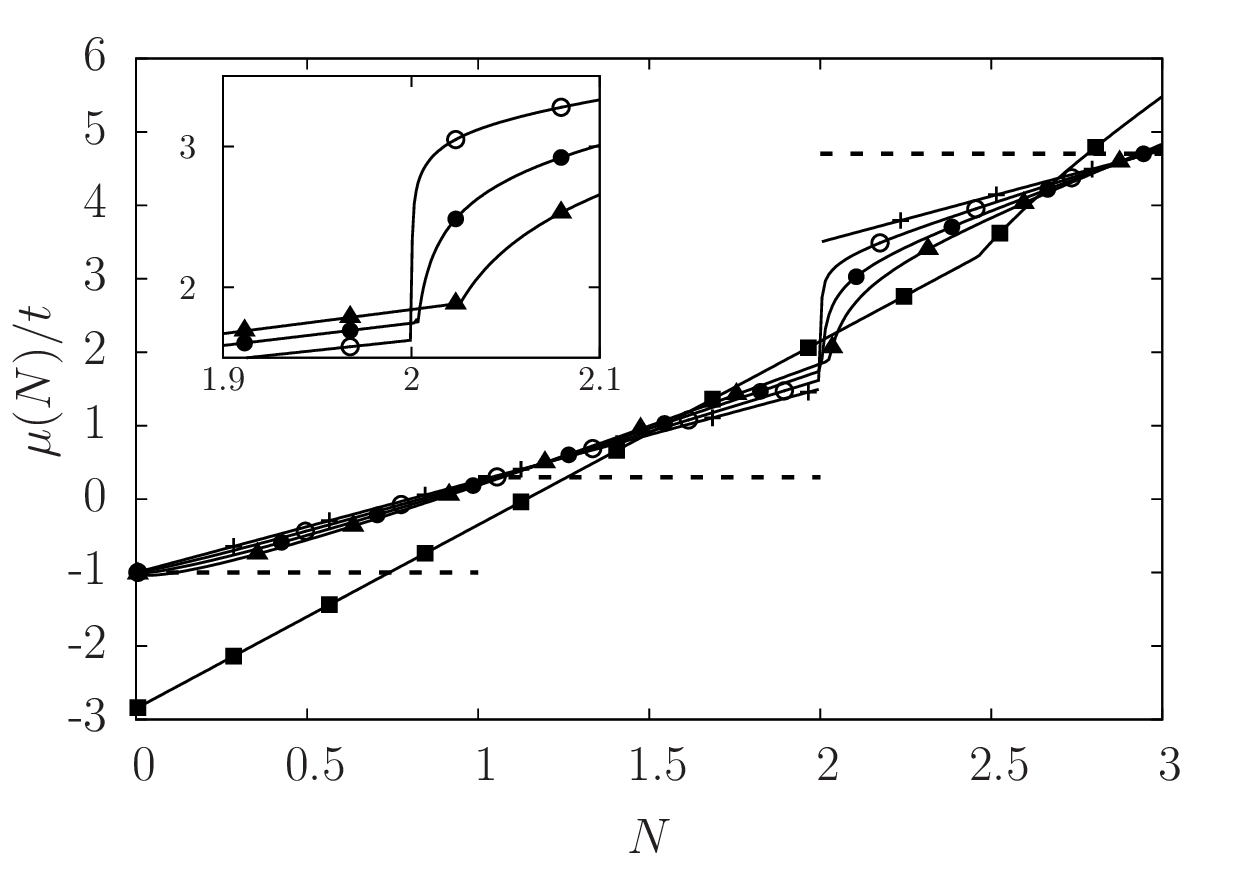}
  \caption{\label{fig:DerDiscon_mu_nonmagn}Chemical potential
      $\mu(N)$ of the Hubbard dimer with $U=5t$ in units of the hopping
      parameter $t$ as function of particle number $N$, when the
      density matrix is restricted to be non-magnetic. 
      The behavior of the power functional with $1/2<\alpha<1$ close
      to half-filling is shown in the inset.  Dashed line: exact solution, crosses:
      Hartree-Fock approximation, open circles: power functional with
      $\alpha=0.95$, filled circles: power functional with
      $\alpha=0.9$, triangles: power functional with $\alpha=0.85$,
      squares: M\"uller functional.  }
\end{figure}

The absence of a true derivative discontinuity using the
  power functional and the tunability of the band gap determined with the
  extrapolation method is not limited to the antiferromagnetic ground
  state. As shown in Fig.~\ref{fig:DerDiscon_mu_nonmagn}, the Hubbard
  dimer behaves qualitatively similar, when the spin polarization is
  suppressed.  In the nonmagnetic calculations, the onset of a finite
  band gap obtained by the extrapolation method is delayed to larger
  power parameters $\alpha$. This finding is analogous to that
  observed for NiO, for which non-magnetic calculations find a
  metallic ground state for $\alpha<0.65$\cite{ShinoharaJCTC_11_4895}, whereas
  non-collinear calculations find an insulating ground state already
  for $\alpha=0.56$\cite{New_J._Phys._17_093038}.

\section{Beyond the dimer}
\label{sec:larger systems}
The question remains whether the findings for the Hubbard dimer
persist in larger systems with more degrees of freedom. This is
  relevant for calculations of more complex systems having large unit
  cells. For this purpose we performed calculations for the power
functional for Hubbard rings and Hubbard chains.

\begin{figure}[htb]
\includegraphics[width=\linewidth,height=!]{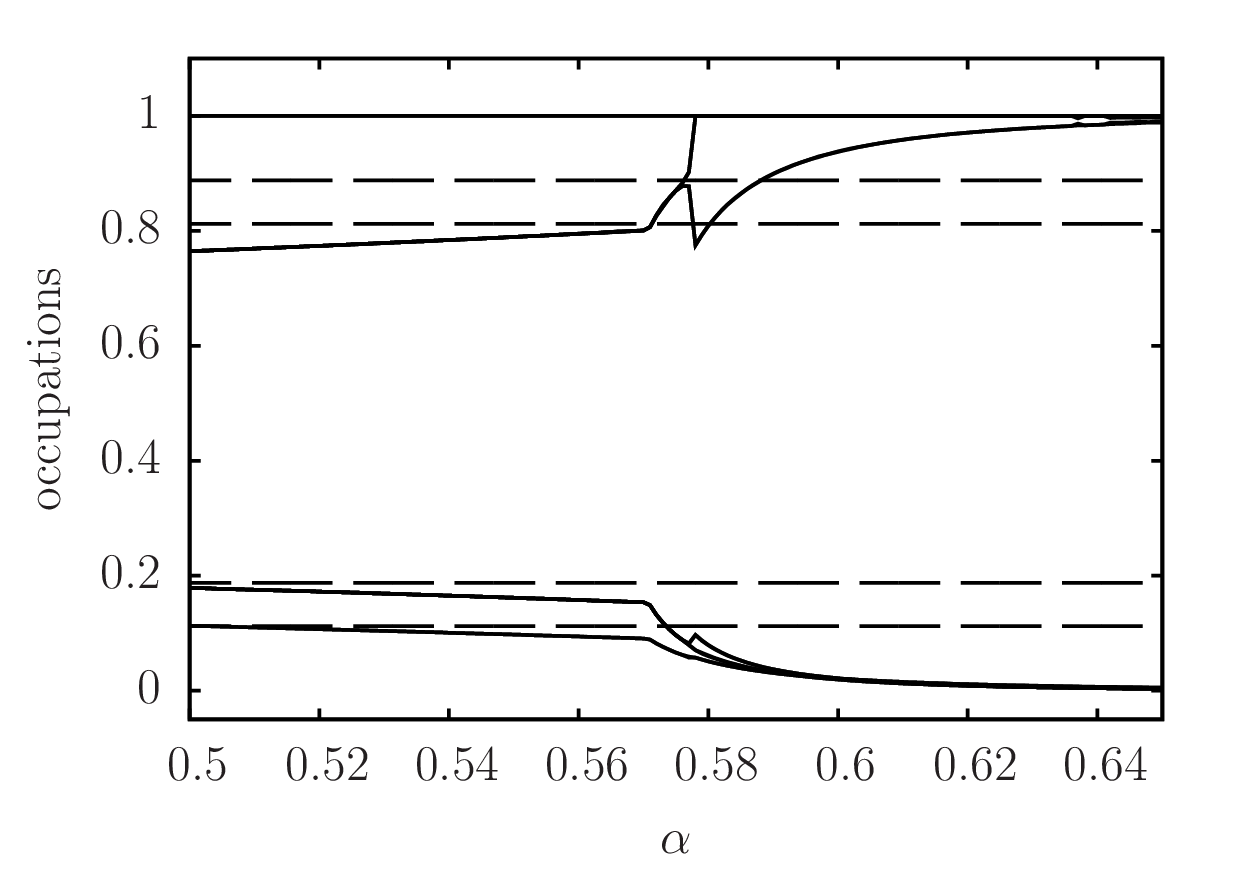}
  \caption{\label{fig:NCHI_12_NPARTICLE_6}Occupations of the
    half-filled six-site Hubbard-ring with $U=5t$ for
    the power functional as function of the parameter $\alpha$ (solid
    lines).  The dashed horizontal lines indicate the occupations of
    the exact many-electron description. Evident are the rather abrupt
    transitions from fractional to integer occupations.}
\end{figure}

Figure~\ref{fig:NCHI_12_NPARTICLE_6} shows the occupation numbers for
a half-filled Hubbard ring at an intermediate interaction strength of
$U=5t$, which like the Hubbard dimer, has an antiferromagnetic ground
state in the Hartree-Fock approximation. For the M\"uller functional
we obtain fractional occupations as for the dimer. While the
fractional occupations deviate from the exact result, their deviation
from integer occupations are of the same order of magnitude as in the
exact solution. The power functional exhibits abrupt transitions to an
antiferromagnetic state around $\alpha_c\approx 0.58$ very analogous
to the Hubbard dimer.

\begin{figure}[htb]
\includegraphics[width=\linewidth,height=!]{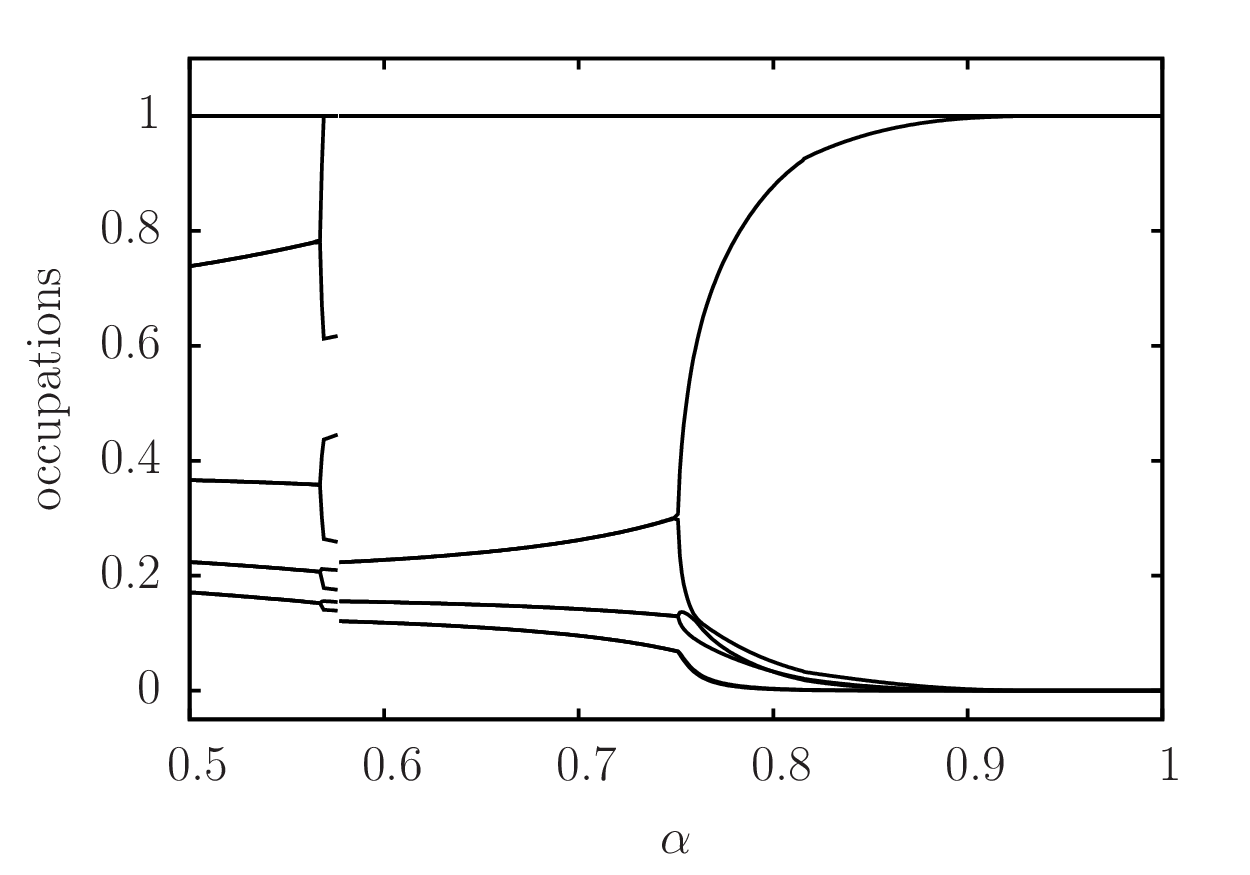}
  \caption{\label{fig:NCHI_12_NPARTICLE_7}Occupations of the six-site
    Hubbard-chain with seven electrons and $U=5t$ for ground states of
    the power functional as function of the parameter $\alpha$
    spanning the range from the M\"uller functional ($\alpha=1/2$) to
    the Hartree-Fock approximation ($\alpha=1$). Evident are the
    rather abrupt transitions from fractional to integer
    occupations.}
\end{figure}

For a six-site Hubbard chain with seven electrons, i.e. one electron
more than half filling, the pattern of transitions is even more
complex: This behavior is shown in
Figure~\ref{fig:NCHI_12_NPARTICLE_7}.  
There are now three transitions: 
\begin{enumerate}
\item A continuous transition between $\alpha\approx0.567$ and
$\alpha\approx0.569$ from the non-magnetic M\"uller ground state to a
state with collinear spins in the pattern $\uparrow \downarrow
\uparrow\uparrow\downarrow\uparrow$, which is only stable in a small
window of parameters. 
\item Around $\alpha\approx 0.576$ there is a non-smooth transition
  to a state with an antiferromagnetic spin-structure, i.e. $\uparrow
  \downarrow \uparrow\downarrow\uparrow\downarrow$.
\item Beyond $\alpha\approx0.75$, the antiferromagnetic structure
  breaks up and evolves into the HF-ground state having
  a spin-structure given by $\uparrow \downarrow
  \uparrow\uparrow\downarrow\uparrow$.  
\end{enumerate}

These examples demonstrate that the power functional can generate a
variety of magnetic states even for simple systems.  

\section{Conclusion}
\label{sec:conclusion}

The popular density-matrix functionals, the M\"uller
functional\cite{mueller84_pl105A_446}, the Hartree-Fock approximation
and the power functional\cite{sharma08_prb78_201103}, which
continuously interpolates between the other two, have been benchmarked
for the Hubbard dimer. The Hartree-Fock approximation is, for the
Hubbard model\cite{gutzwiller63_prl10_159, hubbard63_prsla276_238,
  kanamori63_progtheorphys30_275}, analogous to hybrid density
functionals\cite{becke93_jcp98_1372}, that admix a portion of exact
exchange to the exchange-correlation energy. The local interaction of
the Hubbard model acts analogous to the range
separation\cite{savin88_ijqc34_59,heyd03_jcp118_8207}, which
suppresses the long-ranged Coulomb interaction in the Fock term. In
this respect, the Hartree-Fock approximation also captures the main
effects of the LDA+U method\cite{anisimov91_prb44_943}.

Particular emphasis has been given to left-right correlation, the
dominant correlation effect for bond dissociation, which is not
captured in local density
functionals\cite{cohen12_cr112_289}. Left-right correlation describes
that electrons localize on opposite sites of the dimer. This electron
correlation, which increases with the interaction strength, avoids the
energetic cost of the Coulomb repulsion due to double occupancy of a
site.  In the Hartree-Fock approximation, this left-right correlation
leads to an antiferromagnetic state with a spin-up electron mostly
localized on one side and the spin-down electron on the other. This
so-called broken-symmetry state disagrees with the exact solution,
which is a singlet state, having no local moments, but nevertheless
antiferromagnetic correlations similar to the broken symmetry
state. More importantly, however, the antiferromagnetic transition is
an abrupt one and not a continuous buildup of antiferromagnetic
correlations as in the exact solution. The result is a qualitatively
incorrect shape of the total energy during bond dissociation.

The M\"uller functional\cite{mueller84_pl105A_446} establishes
left-right correlation in a fundamentally different manner: while the
natural orbitals are mostly -- in the Hubbard dimer exactly--
independent of the interaction, the occupations become fractional,
which reflects the creation of electron-hole pairs that screen the
interaction. One of the main successes of the M\"uller functional
besides being able to produce fractional occupations correctly, is
that it captures the continuous nature of the transition to the
left-right-correlated state.

Our calculations avoid any bias and allow for arbitrary non-collinear
spin-polarized states. This strategy shall bring all potential
problems to the surface, that would be present in large scale
electronic structure calculations using these density-matrix
functionals.

Our first observation is that the ground state for the M\"uller
functional, which does not have local moments, is degenerate with a
one dimensional manifold of ferromagnetic states. Thus the dimer has
infinite magnetic susceptibility when described with the M\"uller
functional, in contrast to the vanishing zero-temperature
susceptibility of the exact solution of the Hubbard dimer. This large
magnetic polarizability is likely to cause severe problems in extended
electronic structure calculations.

When turning to the power functional\cite{sharma08_prb78_201103}, we
find that the system behaves analogous to the M\"uller functional for
small interactions, while it exhibits a transition to a
Hartree-Fock-like antiferromagnetic state for large interactions. The
critical interaction, where this transition occurs, drops rapidly with
increasing $\alpha$ from infinity in the M\"uller functional to the
Hartree-Fock value $U_{crit}=2t$.

In the small-interaction regime the system is weakly pinned in the
ferromagnetic state corresponding to the largest moment of the
ground-state manifold of the M\"uller functional. 

Our calculations indicate a major deficiency in the description of
magnetic properties for this class of density-matrix functionals. The
problems persist in modified form also for more general Hamiltonians,
which include off-site Coulomb interactions, and for more extended
systems.

Besides the bond-dissociation problem, we investigated the derivative
discontinuity\cite{perdew82_prl49_1691, cohen12_cr112_289} with
changing the number of electrons. A balanced description of the
electron affinity and ionization potential is essential for a
qualitatively correct description of charge transfer.  We find that
the metal-like behavior of the M\"uller functional persists: The
discontinuity of the exchange-correlation energy even offsets the one
of the kinetic energy. The M\"uller functional describes the Hubbard
dimer with vanishing fundamental gap.

The power functional inherits many of
the problems of the M\"uller functional:  There is no derivative
discontinuity in the entire parameter range of the power functional
except for the Hartree-Fock limit. In the low-interaction regime the
solutions are weakly ferromagnetic. Like the Hartree-Fock
approximation, the power functional exhibits an artificial abrupt
magnetic transition with increasing interaction towards an
antiferromagnetic configuration, albeit at a larger critical interaction.
These states are intrinsically non-collinear.

The absence of any derivative discontinuity also for insulating materials is expected to produce an artificial charge transfer between the constituents of large-scale electronic structure calculations. This cast severe doubt on the performance of such density-matrix functionals for complex systems.

While the power functional lacks a derivative
discontinuity, its chemical potential undergoes a continuous
transition between two linear functions, which has been exploited
to extract a band gap from data obtained further away from the
integer particle number\cite{sharma08_prb78_201103,
EPL_77_(2007)_67003, PhysRevB.78.201103,
PhysRevA.79.022504,Z._Phys._Chem_224_467}.

Our calculations indicate, however, that the band gap obtained
from this extrapolation can be tuned by the free parameter $\alpha$
of the power functional between zero and the Hartree-Fock result. The
band gap opens in non-collinear calculations only when in the antiferromagnetic
regime, while it vanishes in the M\"uller-type regime at low
interactions. The opening of a band gap obtained by the
extrapolation method and its tunability are features that persist in
non-magnetic calculations, while the gap opens at a larger value of
the power parameter than in the magnetic calculation. These problems or
signatures of them can be observed in previous
calculations\cite{PhysRevB.78.201103, Z._Phys._Chem_224_467,
  PhysRevLett.110.116403, New_J._Phys._17_093038,
  ShinoharaJCTC_11_4895}. 

The tunability of the band gap is similar to other methods such
as LDA+U\cite{anisimov91_prb44_943} and hybrid density
functionals\cite{becke93_jcp98_1372}. However, the latter methods
exhibit a true derivative discontinuity and their band gap does not
shrink below the Kohn-Sham band gap, which is analogous to the
non-interacting band gap of the Hubbard dimer.

Approximations for ionization potentials \cite{Pernal200571} and spectral functions\cite{PhysRevLett.110.116403} have been introduced on top of rDMFT. The latter method on the one hand yields spectra that agree well with experimental results for transition metal
oxides\cite{PhysRevLett.110.116403,New_J._Phys._17_093038,ShinoharaJCTC_11_4895}
for particular choices of the power functional parameter. On the
other hand investigations on the Hubbard
dimer\cite{disabatino15_jcp143_24108} suggest caution and claim that
the underlying physics is not correctly treated.  

The problems presented here demonstrate potential fundamental
flaws of the class of density-matrix functionals of this
study.  We hope that this study provides a useful reference point for
the development of new density-matrix functionals. We believe
furthermore that our findings call for new approaches for the
construction of density-matrix functionals that make closer contact to
the many-particle description of the electronic
system\cite{bloechl13_prb88_25139}.

\begin{acknowledgements}
We are deeply saddened by the loss of our dear colleague Thomas
Pruschke.  Financial support from the Deutsche Forschungsgemeinschaft
through FOR1346 (project 9) is gratefully acknowledged.
\end{acknowledgements}


\begin{thebibliography}{56}%
\makeatletter
\providecommand \@ifxundefined [1]{%
 \@ifx{#1\undefined}
}%
\providecommand \@ifnum [1]{%
 \ifnum #1\expandafter \@firstoftwo
 \else \expandafter \@secondoftwo
 \fi
}%
\providecommand \@ifx [1]{%
 \ifx #1\expandafter \@firstoftwo
 \else \expandafter \@secondoftwo
 \fi
}%
\providecommand \natexlab [1]{#1}%
\providecommand \enquote  [1]{``#1''}%
\providecommand \bibnamefont  [1]{#1}%
\providecommand \bibfnamefont [1]{#1}%
\providecommand \citenamefont [1]{#1}%
\providecommand \href@noop [0]{\@secondoftwo}%
\providecommand \href [0]{\begingroup \@sanitize@url \@href}%
\providecommand \@href[1]{\@@startlink{#1}\@@href}%
\providecommand \@@href[1]{\endgroup#1\@@endlink}%
\providecommand \@sanitize@url [0]{\catcode `\\12\catcode `\$12\catcode
  `\&12\catcode `\#12\catcode `\^12\catcode `\_12\catcode `\%12\relax}%
\providecommand \@@startlink[1]{}%
\providecommand \@@endlink[0]{}%
\providecommand \url  [0]{\begingroup\@sanitize@url \@url }%
\providecommand \@url [1]{\endgroup\@href {#1}{\urlprefix }}%
\providecommand \urlprefix  [0]{URL }%
\providecommand \Eprint [0]{\href }%
\providecommand \doibase [0]{http://dx.doi.org/}%
\providecommand \selectlanguage [0]{\@gobble}%
\providecommand \bibinfo  [0]{\@secondoftwo}%
\providecommand \bibfield  [0]{\@secondoftwo}%
\providecommand \translation [1]{[#1]}%
\providecommand \BibitemOpen [0]{}%
\providecommand \bibitemStop [0]{}%
\providecommand \bibitemNoStop [0]{.\EOS\space}%
\providecommand \EOS [0]{\spacefactor3000\relax}%
\providecommand \BibitemShut  [1]{\csname bibitem#1\endcsname}%
\let\auto@bib@innerbib\@empty
\bibitem [{\citenamefont {Hohenberg}\ and\ \citenamefont
  {Kohn}(1964)}]{hohenberg64_pr136_B864}%
  \BibitemOpen
  \bibfield  {author} {\bibinfo {author} {\bibfnamefont {P.}~\bibnamefont
  {Hohenberg}}\ and\ \bibinfo {author} {\bibfnamefont {W.}~\bibnamefont
  {Kohn}},\ }\href {\doibase 10.1103/PhysRev.136.B864} {\bibfield  {journal}
  {\bibinfo  {journal} {Phys. Rev.}\ }\textbf {\bibinfo {volume} {136}},\
  \bibinfo {pages} {B864} (\bibinfo {year} {1964})}\BibitemShut {NoStop}%
\bibitem [{\citenamefont {Kohn}\ and\ \citenamefont
  {Sham}(1965)}]{kohn65_pr140_1133}%
  \BibitemOpen
  \bibfield  {author} {\bibinfo {author} {\bibfnamefont {W.}~\bibnamefont
  {Kohn}}\ and\ \bibinfo {author} {\bibfnamefont {L.~J.}\ \bibnamefont
  {Sham}},\ }\href {\doibase 10.1103/PhysRev.140.A1133} {\bibfield  {journal}
  {\bibinfo  {journal} {Phys. Rev.}\ }\textbf {\bibinfo {volume} {140}},\
  \bibinfo {pages} {A1133} (\bibinfo {year} {1965})}\BibitemShut {NoStop}%
\bibitem [{\citenamefont {Cramer}\ and\ \citenamefont
  {Truhlar}(2009)}]{cramer09_pccp11_10757}%
  \BibitemOpen
  \bibfield  {author} {\bibinfo {author} {\bibfnamefont {C.~J.}\ \bibnamefont
  {Cramer}}\ and\ \bibinfo {author} {\bibfnamefont {D.~G.}\ \bibnamefont
  {Truhlar}},\ }\href {\doibase 10.1039/B907148B} {\bibfield  {journal}
  {\bibinfo  {journal} {Phys. Chem. Chem. Phys.}\ }\textbf {\bibinfo {volume}
  {11}},\ \bibinfo {pages} {10757} (\bibinfo {year} {2009})}\BibitemShut
  {NoStop}%
\bibitem [{\citenamefont {von Barth}(2004)}]{vonbarth04_physicascripta109_9}%
  \BibitemOpen
  \bibfield  {author} {\bibinfo {author} {\bibfnamefont {U.}~\bibnamefont {von
  Barth}},\ }\href {http://stacks.iop.org/1402-4896/2004/i=T109/a=001}
  {\bibfield  {journal} {\bibinfo  {journal} {Physica Scripta}\ }\textbf
  {\bibinfo {volume} {2004}},\ \bibinfo {pages} {9} (\bibinfo {year}
  {2004})}\BibitemShut {NoStop}%
\bibitem [{\citenamefont {Cohen}\ \emph {et~al.}(2012)\citenamefont {Cohen},
  \citenamefont {Mori-S\'anchez},\ and\ \citenamefont
  {Yang}}]{cohen12_cr112_289}%
  \BibitemOpen
  \bibfield  {author} {\bibinfo {author} {\bibfnamefont {A.~J.}\ \bibnamefont
  {Cohen}}, \bibinfo {author} {\bibfnamefont {P.}~\bibnamefont
  {Mori-S\'anchez}}, \ and\ \bibinfo {author} {\bibfnamefont {W.}~\bibnamefont
  {Yang}},\ }\href {\doibase 10.1021/cr200107z} {\bibfield  {journal} {\bibinfo
   {journal} {Chem. Rev.}\ }\textbf {\bibinfo {volume} {112}},\ \bibinfo
  {pages} {289} (\bibinfo {year} {2012})}\BibitemShut {NoStop}%
\bibitem [{\citenamefont {Terakura}\ \emph {et~al.}(1984)\citenamefont
  {Terakura}, \citenamefont {Oguchi}, \citenamefont {Williams},\ and\
  \citenamefont {K\"ubler}}]{terakura84_prb30_4734}%
  \BibitemOpen
  \bibfield  {author} {\bibinfo {author} {\bibfnamefont {K.}~\bibnamefont
  {Terakura}}, \bibinfo {author} {\bibfnamefont {T.}~\bibnamefont {Oguchi}},
  \bibinfo {author} {\bibfnamefont {A.~R.}\ \bibnamefont {Williams}}, \ and\
  \bibinfo {author} {\bibfnamefont {J.}~\bibnamefont {K\"ubler}},\ }\href
  {\doibase 10.1103/PhysRevB.30.4734} {\bibfield  {journal} {\bibinfo
  {journal} {Phys. Rev. B}\ }\textbf {\bibinfo {volume} {30}},\ \bibinfo
  {pages} {4734} (\bibinfo {year} {1984})}\BibitemShut {NoStop}%
\bibitem [{\citenamefont {Anisimov}\ \emph {et~al.}(1991)\citenamefont
  {Anisimov}, \citenamefont {Zaanen},\ and\ \citenamefont
  {Andersen}}]{anisimov91_prb44_943}%
  \BibitemOpen
  \bibfield  {author} {\bibinfo {author} {\bibfnamefont {V.~I.}\ \bibnamefont
  {Anisimov}}, \bibinfo {author} {\bibfnamefont {J.}~\bibnamefont {Zaanen}}, \
  and\ \bibinfo {author} {\bibfnamefont {O.~K.}\ \bibnamefont {Andersen}},\
  }\href {\doibase 10.1103/PhysRevB.44.943} {\bibfield  {journal} {\bibinfo
  {journal} {Phys. Rev. B}\ }\textbf {\bibinfo {volume} {44}},\ \bibinfo
  {pages} {943} (\bibinfo {year} {1991})}\BibitemShut {NoStop}%
\bibitem [{\citenamefont {Georges}\ \emph {et~al.}(1996)\citenamefont
  {Georges}, \citenamefont {Kotliar}, \citenamefont {Krauth},\ and\
  \citenamefont {Rozenberg}}]{georges96_rmp68_13}%
  \BibitemOpen
  \bibfield  {author} {\bibinfo {author} {\bibfnamefont {A.}~\bibnamefont
  {Georges}}, \bibinfo {author} {\bibfnamefont {G.}~\bibnamefont {Kotliar}},
  \bibinfo {author} {\bibfnamefont {W.}~\bibnamefont {Krauth}}, \ and\ \bibinfo
  {author} {\bibfnamefont {M.~J.}\ \bibnamefont {Rozenberg}},\ }\href {\doibase
  doi:10.1103/RevModPhys.68.13} {\bibfield  {journal} {\bibinfo  {journal}
  {Rev. Mod. Phys.}\ }\textbf {\bibinfo {volume} {68}},\ \bibinfo {pages} {13}
  (\bibinfo {year} {1996})}\BibitemShut {NoStop}%
\bibitem [{\citenamefont {Held}(2007)}]{held07_advp56_829}%
  \BibitemOpen
  \bibfield  {author} {\bibinfo {author} {\bibfnamefont {K.}~\bibnamefont
  {Held}},\ }\href {\doibase doi:10.1080/00018730701619647} {\bibfield
  {journal} {\bibinfo  {journal} {Adv. Phys.}\ }\textbf {\bibinfo {volume}
  {56}},\ \bibinfo {pages} {829} (\bibinfo {year} {2007})}\BibitemShut
  {NoStop}%
\bibitem [{\citenamefont {Kotliar}\ \emph {et~al.}(2006)\citenamefont
  {Kotliar}, \citenamefont {Savrasov}, \citenamefont {Haule}, \citenamefont
  {Oudovenko}, \citenamefont {Parcollet},\ and\ \citenamefont
  {Marianetti}}]{kotliar06_rmp78_865}%
  \BibitemOpen
  \bibfield  {author} {\bibinfo {author} {\bibfnamefont {G.}~\bibnamefont
  {Kotliar}}, \bibinfo {author} {\bibfnamefont {S.~Y.}\ \bibnamefont
  {Savrasov}}, \bibinfo {author} {\bibfnamefont {K.}~\bibnamefont {Haule}},
  \bibinfo {author} {\bibfnamefont {V.~S.}\ \bibnamefont {Oudovenko}}, \bibinfo
  {author} {\bibfnamefont {O.}~\bibnamefont {Parcollet}}, \ and\ \bibinfo
  {author} {\bibfnamefont {C.~A.}\ \bibnamefont {Marianetti}},\ }\href
  {\doibase 10.1103/RevModPhys.78.865} {\bibfield  {journal} {\bibinfo
  {journal} {Rev. Mod. Phys.}\ }\textbf {\bibinfo {volume} {78}},\ \bibinfo
  {pages} {865} (\bibinfo {year} {2006})}\BibitemShut {NoStop}%
\bibitem [{\citenamefont {B\"unemann}\ \emph {et~al.}(1998)\citenamefont
  {B\"unemann}, \citenamefont {Weber},\ and\ \citenamefont
  {Gebhard}}]{buenemann98_prb57_6896}%
  \BibitemOpen
  \bibfield  {author} {\bibinfo {author} {\bibfnamefont {J.}~\bibnamefont
  {B\"unemann}}, \bibinfo {author} {\bibfnamefont {W.}~\bibnamefont {Weber}}, \
  and\ \bibinfo {author} {\bibfnamefont {F.}~\bibnamefont {Gebhard}},\ }\href
  {\doibase 10.1103/PhysRevB.57.6896} {\bibfield  {journal} {\bibinfo
  {journal} {Phys. Rev. B}\ }\textbf {\bibinfo {volume} {57}},\ \bibinfo
  {pages} {6896} (\bibinfo {year} {1998})}\BibitemShut {NoStop}%
\bibitem [{\citenamefont {Schickling}\ \emph {et~al.}(2014)\citenamefont
  {Schickling}, \citenamefont {B\"ünemann}, \citenamefont {Gebhard},\ and\
  \citenamefont {Weber}}]{schickling14_njp16_83034}%
  \BibitemOpen
  \bibfield  {author} {\bibinfo {author} {\bibfnamefont {T.}~\bibnamefont
  {Schickling}}, \bibinfo {author} {\bibfnamefont {J.}~\bibnamefont
  {B\"ünemann}}, \bibinfo {author} {\bibfnamefont {F.}~\bibnamefont
  {Gebhard}}, \ and\ \bibinfo {author} {\bibfnamefont {W.}~\bibnamefont
  {Weber}},\ }\href {\doibase doi:10.1088/1367-2630/16/9/093034} {\bibfield
  {journal} {\bibinfo  {journal} {New J. Phys.}\ }\textbf {\bibinfo {volume}
  {16}},\ \bibinfo {pages} {83034} (\bibinfo {year} {2014})}\BibitemShut
  {NoStop}%
\bibitem [{\citenamefont {Wang}\ \emph {et~al.}(2008)\citenamefont {Wang},
  \citenamefont {Dai},\ and\ \citenamefont {Fang}}]{wang08_prl101_66403}%
  \BibitemOpen
  \bibfield  {author} {\bibinfo {author} {\bibfnamefont {G.-T.}\ \bibnamefont
  {Wang}}, \bibinfo {author} {\bibfnamefont {X.}~\bibnamefont {Dai}}, \ and\
  \bibinfo {author} {\bibfnamefont {Z.}~\bibnamefont {Fang}},\ }\href {\doibase
  10.1103/PhysRevLett.101.066403} {\bibfield  {journal} {\bibinfo  {journal}
  {Phys. Rev. Lett.}\ }\textbf {\bibinfo {volume} {101}},\ \bibinfo {pages}
  {066403} (\bibinfo {year} {2008})}\BibitemShut {NoStop}%
\bibitem [{\citenamefont {Gutzwiller}(1963)}]{gutzwiller63_prl10_159}%
  \BibitemOpen
  \bibfield  {author} {\bibinfo {author} {\bibfnamefont {M.~C.}\ \bibnamefont
  {Gutzwiller}},\ }\href {\doibase 10.1103/PhysRevLett.10.159} {\bibfield
  {journal} {\bibinfo  {journal} {Phys. Rev. Lett.}\ }\textbf {\bibinfo
  {volume} {10}},\ \bibinfo {pages} {159} (\bibinfo {year} {1963})}\BibitemShut
  {NoStop}%
\bibitem [{\citenamefont {Hubbard}(1963)}]{hubbard63_prsla276_238}%
  \BibitemOpen
  \bibfield  {author} {\bibinfo {author} {\bibfnamefont {J.}~\bibnamefont
  {Hubbard}},\ }\href {\doibase 10.1098/rspa.1963.0204} {\bibfield  {journal}
  {\bibinfo  {journal} {Proc. R. Soc. Lond. A}\ }\textbf {\bibinfo {volume}
  {276}},\ \bibinfo {pages} {238} (\bibinfo {year} {1963})}\BibitemShut
  {NoStop}%
\bibitem [{\citenamefont {Kanamori}(1963)}]{kanamori63_progtheorphys30_275}%
  \BibitemOpen
  \bibfield  {author} {\bibinfo {author} {\bibfnamefont {J.}~\bibnamefont
  {Kanamori}},\ }\href {\doibase doi:10.1143/PTP.30.275} {\bibfield  {journal}
  {\bibinfo  {journal} {Prog. Theor. Phys.}\ }\textbf {\bibinfo {volume}
  {30}},\ \bibinfo {pages} {275} (\bibinfo {year} {1963})}\BibitemShut
  {NoStop}%
\bibitem [{\citenamefont {Gilbert}(1975)}]{gilbert75_prb12_2111}%
  \BibitemOpen
  \bibfield  {author} {\bibinfo {author} {\bibfnamefont {T.~L.}\ \bibnamefont
  {Gilbert}},\ }\href {\doibase 10.1103/PhysRevB.12.2111} {\bibfield  {journal}
  {\bibinfo  {journal} {Phys. Rev. B}\ }\textbf {\bibinfo {volume} {12}},\
  \bibinfo {pages} {2111} (\bibinfo {year} {1975})}\BibitemShut {NoStop}%
\bibitem [{\citenamefont {Levy}(1979)}]{levy79_pnas76_6062}%
  \BibitemOpen
  \bibfield  {author} {\bibinfo {author} {\bibfnamefont {M.}~\bibnamefont
  {Levy}},\ }\href {http://www.pnas.org/content/76/12/6062.abstract} {\bibfield
   {journal} {\bibinfo  {journal} {Proc. Nat'l Acad. Sci. USA}\ }\textbf
  {\bibinfo {volume} {76}},\ \bibinfo {pages} {6062} (\bibinfo {year}
  {1979})}\BibitemShut {NoStop}%
\bibitem [{\citenamefont {Bl\"ochl}\ \emph {et~al.}(2011)\citenamefont
  {Bl\"ochl}, \citenamefont {Walther},\ and\ \citenamefont
  {Pruschke}}]{bloechl11_prb84_205101}%
  \BibitemOpen
  \bibfield  {author} {\bibinfo {author} {\bibfnamefont {P.~E.}\ \bibnamefont
  {Bl\"ochl}}, \bibinfo {author} {\bibfnamefont {C.~F.~J.}\ \bibnamefont
  {Walther}}, \ and\ \bibinfo {author} {\bibfnamefont {T.}~\bibnamefont
  {Pruschke}},\ }\href {\doibase 10.1103/PhysRevB.84.205101} {\bibfield
  {journal} {\bibinfo  {journal} {Phys. Rev. B}\ }\textbf {\bibinfo {volume}
  {84}},\ \bibinfo {pages} {205101} (\bibinfo {year} {2011})}\BibitemShut
  {NoStop}%
\bibitem [{\citenamefont {Bl\"ochl}\ \emph {et~al.}(2013)\citenamefont
  {Bl\"ochl}, \citenamefont {Pruschke},\ and\ \citenamefont
  {Potthoff}}]{bloechl13_prb88_25139}%
  \BibitemOpen
  \bibfield  {author} {\bibinfo {author} {\bibfnamefont {P.~E.}\ \bibnamefont
  {Bl\"ochl}}, \bibinfo {author} {\bibfnamefont {T.}~\bibnamefont {Pruschke}},
  \ and\ \bibinfo {author} {\bibfnamefont {M.}~\bibnamefont {Potthoff}},\
  }\href {\doibase 10.1103/PhysRevB.88.205139} {\bibfield  {journal} {\bibinfo
  {journal} {Phys. Rev. B}\ }\textbf {\bibinfo {volume} {88}},\ \bibinfo
  {pages} {205139} (\bibinfo {year} {2013})}\BibitemShut {NoStop}%
\bibitem [{\citenamefont {Luttinger}\ and\ \citenamefont
  {Ward}(1960)}]{luttinger60_pr118_1417}%
  \BibitemOpen
  \bibfield  {author} {\bibinfo {author} {\bibfnamefont {J.~M.}\ \bibnamefont
  {Luttinger}}\ and\ \bibinfo {author} {\bibfnamefont {J.~C.}\ \bibnamefont
  {Ward}},\ }\href {\doibase 10.1103/PhysRev.118.1417} {\bibfield  {journal}
  {\bibinfo  {journal} {Phys. Rev.}\ }\textbf {\bibinfo {volume} {118}},\
  \bibinfo {pages} {1417} (\bibinfo {year} {1960})}\BibitemShut {NoStop}%
\bibitem [{\citenamefont {M\"uller}(1984)}]{mueller84_pl105A_446}%
  \BibitemOpen
  \bibfield  {author} {\bibinfo {author} {\bibfnamefont {A.~M.~K.}\
  \bibnamefont {M\"uller}},\ }\href {\doibase doi:10.1016/0375-9601(84)91034-X}
  {\bibfield  {journal} {\bibinfo  {journal} {Phys. Lett.}\ }\textbf {\bibinfo
  {volume} {105}},\ \bibinfo {pages} {446} (\bibinfo {year}
  {1984})}\BibitemShut {NoStop}%
\bibitem [{\citenamefont {Goedecker}\ and\ \citenamefont
  {Umrigar}(1998)}]{goedecker98_prl81_866}%
  \BibitemOpen
  \bibfield  {author} {\bibinfo {author} {\bibfnamefont {S.}~\bibnamefont
  {Goedecker}}\ and\ \bibinfo {author} {\bibfnamefont {C.~J.}\ \bibnamefont
  {Umrigar}},\ }\href {\doibase 10.1103/PhysRevLett.81.866} {\bibfield
  {journal} {\bibinfo  {journal} {Phys. Rev. Lett.}\ }\textbf {\bibinfo
  {volume} {81}},\ \bibinfo {pages} {866} (\bibinfo {year} {1998})}\BibitemShut
  {NoStop}%
\bibitem [{\citenamefont {Gritsenko}\ \emph {et~al.}(2005)\citenamefont
  {Gritsenko}, \citenamefont {Pernal},\ and\ \citenamefont
  {Baerends}}]{gritsenko05_jcp122_204102}%
  \BibitemOpen
  \bibfield  {author} {\bibinfo {author} {\bibfnamefont {O.}~\bibnamefont
  {Gritsenko}}, \bibinfo {author} {\bibfnamefont {K.}~\bibnamefont {Pernal}}, \
  and\ \bibinfo {author} {\bibfnamefont {E.~J.}\ \bibnamefont {Baerends}},\
  }\href {\doibase doi:10.1063/1.1906203} {\bibfield  {journal} {\bibinfo
  {journal} {J. Chem. Phys.}\ }\textbf {\bibinfo {volume} {122}},\ \bibinfo
  {eid} {204102} (\bibinfo {year} {2005})}\BibitemShut {NoStop}%
\bibitem [{\citenamefont {Sharma}\ \emph
  {et~al.}(2008{\natexlab{a}})\citenamefont {Sharma}, \citenamefont {Dewhurst},
  \citenamefont {Lathiotakis},\ and\ \citenamefont
  {Gross}}]{sharma08_prb78_201103}%
  \BibitemOpen
  \bibfield  {author} {\bibinfo {author} {\bibfnamefont {S.}~\bibnamefont
  {Sharma}}, \bibinfo {author} {\bibfnamefont {J.~K.}\ \bibnamefont
  {Dewhurst}}, \bibinfo {author} {\bibfnamefont {N.~N.}\ \bibnamefont
  {Lathiotakis}}, \ and\ \bibinfo {author} {\bibfnamefont {E.~K.~U.}\
  \bibnamefont {Gross}},\ }\href {\doibase 10.1103/PhysRevB.78.201103}
  {\bibfield  {journal} {\bibinfo  {journal} {Phys. Rev. B}\ }\textbf {\bibinfo
  {volume} {78}},\ \bibinfo {pages} {201103} (\bibinfo {year}
  {2008}{\natexlab{a}})}\BibitemShut {NoStop}%
\bibitem [{\citenamefont {Marques}\ and\ \citenamefont
  {Lathiotakis}(2008)}]{marques08_pra77_32509}%
  \BibitemOpen
  \bibfield  {author} {\bibinfo {author} {\bibfnamefont {M.~A.~L.}\
  \bibnamefont {Marques}}\ and\ \bibinfo {author} {\bibfnamefont {N.~N.}\
  \bibnamefont {Lathiotakis}},\ }\href {\doibase 10.1103/PhysRevA.77.032509}
  {\bibfield  {journal} {\bibinfo  {journal} {Phys. Rev. A}\ }\textbf {\bibinfo
  {volume} {77}},\ \bibinfo {pages} {032509} (\bibinfo {year}
  {2008})}\BibitemShut {NoStop}%
\bibitem [{\citenamefont {Benavides-Riveros}\ and\ \citenamefont
  {V\'arilly}(2012)}]{benavidesriveros12_epjd66_274}%
  \BibitemOpen
  \bibfield  {author} {\bibinfo {author} {\bibfnamefont {C.}~\bibnamefont
  {Benavides-Riveros}}\ and\ \bibinfo {author} {\bibfnamefont {J.}~\bibnamefont
  {V\'arilly}},\ }\href {http://dx.doi.org/10.1140/epjd/e2012-30442-4}
  {\bibfield  {journal} {\bibinfo  {journal} {Eur. Phys. J. D}\ }\textbf
  {\bibinfo {volume} {66}},\ \bibinfo {eid} {274} (\bibinfo {year}
  {2012})}\BibitemShut {NoStop}%
\bibitem [{\citenamefont {Lathiotakis}\ \emph {et~al.}(2009)\citenamefont
  {Lathiotakis}, \citenamefont {Sharma}, \citenamefont {Dewhurst},
  \citenamefont {Eich}, \citenamefont {Marques},\ and\ \citenamefont
  {Gross}}]{lathiotakis09_pra79_40501}%
  \BibitemOpen
  \bibfield  {author} {\bibinfo {author} {\bibfnamefont {N.~N.}\ \bibnamefont
  {Lathiotakis}}, \bibinfo {author} {\bibfnamefont {S.}~\bibnamefont {Sharma}},
  \bibinfo {author} {\bibfnamefont {J.~K.}\ \bibnamefont {Dewhurst}}, \bibinfo
  {author} {\bibfnamefont {F.~G.}\ \bibnamefont {Eich}}, \bibinfo {author}
  {\bibfnamefont {M.~A.~L.}\ \bibnamefont {Marques}}, \ and\ \bibinfo {author}
  {\bibfnamefont {E.~K.~U.}\ \bibnamefont {Gross}},\ }\href {\doibase
  10.1103/PhysRevA.79.040501} {\bibfield  {journal} {\bibinfo  {journal} {Phys.
  Rev. A}\ }\textbf {\bibinfo {volume} {79}},\ \bibinfo {pages} {040501}
  (\bibinfo {year} {2009})}\BibitemShut {NoStop}%
\bibitem [{\citenamefont {Olsen}\ and\ \citenamefont
  {Thygesen}(2014)}]{olsen14_jcp140_164116}%
  \BibitemOpen
  \bibfield  {author} {\bibinfo {author} {\bibfnamefont {T.}~\bibnamefont
  {Olsen}}\ and\ \bibinfo {author} {\bibfnamefont {K.~S.}\ \bibnamefont
  {Thygesen}},\ }\href
  {http://scitation.aip.org/content/aip/journal/jcp/140/16/10.1063/1.4871875}
  {\bibfield  {journal} {\bibinfo  {journal} {J. Chem. Phys.}\ }\textbf
  {\bibinfo {volume} {140}},\ \bibinfo {eid} {164116} (\bibinfo {year}
  {2014})}\BibitemShut {NoStop}%
\bibitem [{\citenamefont {di~Sabatino}\ \emph {et~al.}(2015)\citenamefont
  {di~Sabatino}, \citenamefont {Berger}, \citenamefont {Reining},\ and\
  \citenamefont {Romaniello}}]{disabatino15_jcp143_24108}%
  \BibitemOpen
  \bibfield  {author} {\bibinfo {author} {\bibfnamefont {S.}~\bibnamefont
  {di~Sabatino}}, \bibinfo {author} {\bibfnamefont {J.~A.}\ \bibnamefont
  {Berger}}, \bibinfo {author} {\bibfnamefont {L.}~\bibnamefont {Reining}}, \
  and\ \bibinfo {author} {\bibfnamefont {P.}~\bibnamefont {Romaniello}},\
  }\href {\doibase doi:10.1063/1.4926327} {\bibfield  {journal} {\bibinfo
  {journal} {J. Chem. Phys.}\ }\textbf {\bibinfo {volume} {143}},\ \bibinfo
  {eid} {024108} (\bibinfo {year} {2015})}\BibitemShut {NoStop}%
\bibitem [{\citenamefont {Sharma}\ \emph {et~al.}(2013)\citenamefont {Sharma},
  \citenamefont {Dewhurst}, \citenamefont {Shallcross},\ and\ \citenamefont
  {Gross}}]{PhysRevLett.110.116403}%
  \BibitemOpen
  \bibfield  {author} {\bibinfo {author} {\bibfnamefont {S.}~\bibnamefont
  {Sharma}}, \bibinfo {author} {\bibfnamefont {J.~K.}\ \bibnamefont
  {Dewhurst}}, \bibinfo {author} {\bibfnamefont {S.}~\bibnamefont
  {Shallcross}}, \ and\ \bibinfo {author} {\bibfnamefont {E.~K.~U.}\
  \bibnamefont {Gross}},\ }\href {\doibase 10.1103/PhysRevLett.110.116403}
  {\bibfield  {journal} {\bibinfo  {journal} {Phys. Rev. Lett.}\ }\textbf
  {\bibinfo {volume} {110}},\ \bibinfo {pages} {116403} (\bibinfo {year}
  {2013})}\BibitemShut {NoStop}%
\bibitem [{\citenamefont {Carrascal}\ \emph {et~al.}(2015)\citenamefont
  {Carrascal}, \citenamefont {Ferrer}, \citenamefont {Smith},\ and\
  \citenamefont {Burke}}]{0953-8984-27-39-393001}%
  \BibitemOpen
  \bibfield  {author} {\bibinfo {author} {\bibfnamefont {D.~J.}\ \bibnamefont
  {Carrascal}}, \bibinfo {author} {\bibfnamefont {J.}~\bibnamefont {Ferrer}},
  \bibinfo {author} {\bibfnamefont {J.~C.}\ \bibnamefont {Smith}}, \ and\
  \bibinfo {author} {\bibfnamefont {K.}~\bibnamefont {Burke}},\ }\href
  {http://stacks.iop.org/0953-8984/27/i=39/a=393001} {\bibfield  {journal}
  {\bibinfo  {journal} {Journal of Physics: Condensed Matter}\ }\textbf
  {\bibinfo {volume} {27}},\ \bibinfo {pages} {393001} (\bibinfo {year}
  {2015})}\BibitemShut {NoStop}%
\bibitem [{\citenamefont {L\"owdin}(1955)}]{loewdin55_pr97_1474}%
  \BibitemOpen
  \bibfield  {author} {\bibinfo {author} {\bibfnamefont {P.-O.}\ \bibnamefont
  {L\"owdin}},\ }\href {\doibase 10.1103/PhysRev.97.1474} {\bibfield  {journal}
  {\bibinfo  {journal} {Phys. Rev.}\ }\textbf {\bibinfo {volume} {97}},\
  \bibinfo {pages} {1474} (\bibinfo {year} {1955})}\BibitemShut {NoStop}%
\bibitem [{\citenamefont {Coleman}(1963)}]{coleman63_rmp35_668}%
  \BibitemOpen
  \bibfield  {author} {\bibinfo {author} {\bibfnamefont {A.}~\bibnamefont
  {Coleman}},\ }\href {\doibase 10.1103/RevModPhys.35.668} {\bibfield
  {journal} {\bibinfo  {journal} {Rev. Mod. Phys.}\ }\textbf {\bibinfo {volume}
  {35}},\ \bibinfo {pages} {668} (\bibinfo {year} {1963})}\BibitemShut
  {NoStop}%
\bibitem [{\citenamefont {Lieb}(1983)}]{lieb83_ijqc24_243}%
  \BibitemOpen
  \bibfield  {author} {\bibinfo {author} {\bibfnamefont {E.~H.}\ \bibnamefont
  {Lieb}},\ }\href {\doibase 10.1002/qua.560240302} {\bibfield  {journal}
  {\bibinfo  {journal} {Int. J. Quantum Chem.}\ }\textbf {\bibinfo {volume}
  {24}},\ \bibinfo {pages} {243} (\bibinfo {year} {1983})}\BibitemShut
  {NoStop}%
\bibitem [{\citenamefont {Baldsiefen}\ \emph {et~al.}(2015)\citenamefont
  {Baldsiefen}, \citenamefont {Cangi},\ and\ \citenamefont
  {Gross}}]{PhysRevA.92.052514}%
  \BibitemOpen
  \bibfield  {author} {\bibinfo {author} {\bibfnamefont {T.}~\bibnamefont
  {Baldsiefen}}, \bibinfo {author} {\bibfnamefont {A.}~\bibnamefont {Cangi}}, \
  and\ \bibinfo {author} {\bibfnamefont {E.~K.~U.}\ \bibnamefont {Gross}},\
  }\href {\doibase 10.1103/PhysRevA.92.052514} {\bibfield  {journal} {\bibinfo
  {journal} {Phys. Rev. A}\ }\textbf {\bibinfo {volume} {92}},\ \bibinfo
  {pages} {052514} (\bibinfo {year} {2015})}\BibitemShut {NoStop}%
\bibitem [{\citenamefont {Perdew}\ \emph {et~al.}(1996)\citenamefont {Perdew},
  \citenamefont {Burke},\ and\ \citenamefont {Wang}}]{perdew96_prb54_16533}%
  \BibitemOpen
  \bibfield  {author} {\bibinfo {author} {\bibfnamefont {J.~P.}\ \bibnamefont
  {Perdew}}, \bibinfo {author} {\bibfnamefont {K.}~\bibnamefont {Burke}}, \
  and\ \bibinfo {author} {\bibfnamefont {Y.}~\bibnamefont {Wang}},\ }\href
  {\doibase 10.1103/PhysRevB.54.16533} {\bibfield  {journal} {\bibinfo
  {journal} {Phys. Rev. B}\ }\textbf {\bibinfo {volume} {54}},\ \bibinfo
  {pages} {16533} (\bibinfo {year} {1996})}\BibitemShut {NoStop}%
\bibitem [{\citenamefont {Baerends}(2001)}]{baerends01_prl87_133004}%
  \BibitemOpen
  \bibfield  {author} {\bibinfo {author} {\bibfnamefont {E.~J.}\ \bibnamefont
  {Baerends}},\ }\href {\doibase 10.1103/PhysRevLett.87.133004} {\bibfield
  {journal} {\bibinfo  {journal} {Phys. Rev. Lett}\ }\textbf {\bibinfo {volume}
  {87}},\ \bibinfo {pages} {133004} (\bibinfo {year} {2001})}\BibitemShut
  {NoStop}%
\bibitem [{\citenamefont {Buijse}\ and\ \citenamefont
  {Baerends}(2002)}]{buijse02_molphys100_401}%
  \BibitemOpen
  \bibfield  {author} {\bibinfo {author} {\bibfnamefont {M.}~\bibnamefont
  {Buijse}}\ and\ \bibinfo {author} {\bibfnamefont {E.}~\bibnamefont
  {Baerends}},\ }\href {\doibase 10.1080/00268970110070243} {\bibfield
  {journal} {\bibinfo  {journal} {Mol. Phys.}\ }\textbf {\bibinfo {volume}
  {100}},\ \bibinfo {pages} {401} (\bibinfo {year} {2002})}\BibitemShut
  {NoStop}%
\bibitem [{\citenamefont {Car}\ and\ \citenamefont
  {Parrinello}(1985)}]{car85_prl55_2471}%
  \BibitemOpen
  \bibfield  {author} {\bibinfo {author} {\bibfnamefont {R.}~\bibnamefont
  {Car}}\ and\ \bibinfo {author} {\bibfnamefont {M.}~\bibnamefont
  {Parrinello}},\ }\href {\doibase 10.1103/PhysRevLett.55.2471} {\bibfield
  {journal} {\bibinfo  {journal} {Phys. Rev. Lett}\ }\textbf {\bibinfo {volume}
  {55}},\ \bibinfo {pages} {2471} (\bibinfo {year} {1985})}\BibitemShut
  {NoStop}%
\bibitem [{\citenamefont {Ryckaert}\ \emph {et~al.}(1977)\citenamefont
  {Ryckaert}, \citenamefont {Ciccotti},\ and\ \citenamefont
  {Berendsen}}]{ryckaert77_jcompphys23_327}%
  \BibitemOpen
  \bibfield  {author} {\bibinfo {author} {\bibfnamefont {J.-P.}\ \bibnamefont
  {Ryckaert}}, \bibinfo {author} {\bibfnamefont {G.}~\bibnamefont {Ciccotti}},
  \ and\ \bibinfo {author} {\bibfnamefont {H.~J.~C.}\ \bibnamefont
  {Berendsen}},\ }\href {\doibase doi:10.1016/0021-9991(77)90098-5} {\bibfield
  {journal} {\bibinfo  {journal} {J. Comput. Phys.}\ }\textbf {\bibinfo
  {volume} {23}},\ \bibinfo {pages} {327} (\bibinfo {year} {1977})}\BibitemShut
  {NoStop}%
\bibitem [{\citenamefont {L\"owdin}\ and\ \citenamefont
  {Shull}(1956)}]{PhysRev.101.1730}%
  \BibitemOpen
  \bibfield  {author} {\bibinfo {author} {\bibfnamefont {P.-O.}\ \bibnamefont
  {L\"owdin}}\ and\ \bibinfo {author} {\bibfnamefont {H.}~\bibnamefont
  {Shull}},\ }\href {\doibase 10.1103/PhysRev.101.1730} {\bibfield  {journal}
  {\bibinfo  {journal} {Phys. Rev.}\ }\textbf {\bibinfo {volume} {101}},\
  \bibinfo {pages} {1730} (\bibinfo {year} {1956})}\BibitemShut {NoStop}%
\bibitem [{\citenamefont {Suezaki}(1972)}]{Suezaki1972293}%
  \BibitemOpen
  \bibfield  {author} {\bibinfo {author} {\bibfnamefont {Y.}~\bibnamefont
  {Suezaki}},\ }\href {\doibase doi:10.1016/0375-9601(72)90086-2} {\bibfield
  {journal} {\bibinfo  {journal} {Physics Letters A}\ }\textbf {\bibinfo
  {volume} {38}},\ \bibinfo {pages} {293 } (\bibinfo {year}
  {1972})}\BibitemShut {NoStop}%
\bibitem [{\citenamefont {Bernstein}\ and\ \citenamefont
  {Pincus}(1974)}]{PhysRevB.10.3626}%
  \BibitemOpen
  \bibfield  {author} {\bibinfo {author} {\bibfnamefont {U.}~\bibnamefont
  {Bernstein}}\ and\ \bibinfo {author} {\bibfnamefont {P.}~\bibnamefont
  {Pincus}},\ }\href {\doibase 10.1103/PhysRevB.10.3626} {\bibfield  {journal}
  {\bibinfo  {journal} {Phys. Rev. B}\ }\textbf {\bibinfo {volume} {10}},\
  \bibinfo {pages} {3626} (\bibinfo {year} {1974})}\BibitemShut {NoStop}%
\bibitem [{Note1()}]{Note1}%
  \BibitemOpen
  \bibinfo {note} {For degenerate states and when the electron addition and
  removal are dominated by delocalized states the discontinuity may also vanish
  or become infinitesimally small.}\BibitemShut {Stop}%
\bibitem [{\citenamefont {Helbig}\ \emph {et~al.}(2009)\citenamefont {Helbig},
  \citenamefont {Lathiotakis},\ and\ \citenamefont
  {Gross}}]{PhysRevA.79.022504}%
  \BibitemOpen
  \bibfield  {author} {\bibinfo {author} {\bibfnamefont {N.}~\bibnamefont
  {Helbig}}, \bibinfo {author} {\bibfnamefont {N.~N.}\ \bibnamefont
  {Lathiotakis}}, \ and\ \bibinfo {author} {\bibfnamefont {E.~K.~U.}\
  \bibnamefont {Gross}},\ }\href {\doibase 10.1103/PhysRevA.79.022504}
  {\bibfield  {journal} {\bibinfo  {journal} {Phys. Rev. A}\ }\textbf {\bibinfo
  {volume} {79}},\ \bibinfo {pages} {022504} (\bibinfo {year}
  {2009})}\BibitemShut {NoStop}%
\bibitem [{\citenamefont {Lathiotakis}\ \emph {et~al.}(2010)\citenamefont
  {Lathiotakis}, \citenamefont {Sharma}, \citenamefont {Helbig}, \citenamefont
  {J.K.Dewhurst}, \citenamefont {M.A.L.Marques}, \citenamefont {Eich},
  \citenamefont {Baldsiefen}, \citenamefont {Zacarias},\ and\ \citenamefont
  {Gross}}]{Z._Phys._Chem_224_467}%
  \BibitemOpen
  \bibfield  {author} {\bibinfo {author} {\bibfnamefont {N.~N.}\ \bibnamefont
  {Lathiotakis}}, \bibinfo {author} {\bibfnamefont {S.}~\bibnamefont {Sharma}},
  \bibinfo {author} {\bibfnamefont {N.}~\bibnamefont {Helbig}}, \bibinfo
  {author} {\bibnamefont {J.K.Dewhurst}}, \bibinfo {author} {\bibnamefont
  {M.A.L.Marques}}, \bibinfo {author} {\bibfnamefont {F.}~\bibnamefont {Eich}},
  \bibinfo {author} {\bibfnamefont {T.}~\bibnamefont {Baldsiefen}}, \bibinfo
  {author} {\bibfnamefont {A.}~\bibnamefont {Zacarias}}, \ and\ \bibinfo
  {author} {\bibfnamefont {E.~K.~U.}\ \bibnamefont {Gross}},\ }\href {\doibase
  10.1524/zpch.2010.6118} {\bibfield  {journal} {\bibinfo  {journal}
  {Zeitschrift fuer Physikalische Chemie}\ }\textbf {\bibinfo {volume} {224}},\
  \bibinfo {pages} {467} (\bibinfo {year} {2010})}\BibitemShut {NoStop}%
\bibitem [{\citenamefont {Helbig}\ \emph {et~al.}(2007)\citenamefont {Helbig},
  \citenamefont {Lathiotakis}, \citenamefont {Albrecht},\ and\ \citenamefont
  {Gross}}]{EPL_77_(2007)_67003}%
  \BibitemOpen
  \bibfield  {author} {\bibinfo {author} {\bibfnamefont {N.}~\bibnamefont
  {Helbig}}, \bibinfo {author} {\bibfnamefont {N.~N.}\ \bibnamefont
  {Lathiotakis}}, \bibinfo {author} {\bibfnamefont {M.}~\bibnamefont
  {Albrecht}}, \ and\ \bibinfo {author} {\bibfnamefont {E.~K.~U.}\ \bibnamefont
  {Gross}},\ }\href {http://stacks.iop.org/0295-5075/77/i=6/a=67003} {\bibfield
   {journal} {\bibinfo  {journal} {EPL (Europhysics Letters)}\ }\textbf
  {\bibinfo {volume} {77}},\ \bibinfo {pages} {67003} (\bibinfo {year}
  {2007})}\BibitemShut {NoStop}%
\bibitem [{\citenamefont {Shinohara}\ \emph
  {et~al.}(2015{\natexlab{a}})\citenamefont {Shinohara}, \citenamefont
  {Sharma}, \citenamefont {Shallcross}, \citenamefont {Lathiotakis},\ and\
  \citenamefont {Gross}}]{ShinoharaJCTC_11_4895}%
  \BibitemOpen
  \bibfield  {author} {\bibinfo {author} {\bibfnamefont {Y.}~\bibnamefont
  {Shinohara}}, \bibinfo {author} {\bibfnamefont {S.}~\bibnamefont {Sharma}},
  \bibinfo {author} {\bibfnamefont {S.}~\bibnamefont {Shallcross}}, \bibinfo
  {author} {\bibfnamefont {N.~N.}\ \bibnamefont {Lathiotakis}}, \ and\ \bibinfo
  {author} {\bibfnamefont {E.~K.~U.}\ \bibnamefont {Gross}},\ }\href {\doibase
  10.1021/acs.jctc.5b00661} {\bibfield  {journal} {\bibinfo  {journal} {Journal
  of Chemical Theory and Computation}\ }\textbf {\bibinfo {volume} {11}},\
  \bibinfo {pages} {4895} (\bibinfo {year} {2015}{\natexlab{a}})},\ \bibinfo
  {note} {pMID: 26574277}\BibitemShut {NoStop}%
\bibitem [{\citenamefont {Shinohara}\ \emph
  {et~al.}(2015{\natexlab{b}})\citenamefont {Shinohara}, \citenamefont
  {Sharma}, \citenamefont {Dewhurst}, \citenamefont {Shallcross}, \citenamefont
  {Lathiotakis},\ and\ \citenamefont {Gross}}]{New_J._Phys._17_093038}%
  \BibitemOpen
  \bibfield  {author} {\bibinfo {author} {\bibfnamefont {Y.}~\bibnamefont
  {Shinohara}}, \bibinfo {author} {\bibfnamefont {S.}~\bibnamefont {Sharma}},
  \bibinfo {author} {\bibfnamefont {J.~K.}\ \bibnamefont {Dewhurst}}, \bibinfo
  {author} {\bibfnamefont {S.}~\bibnamefont {Shallcross}}, \bibinfo {author}
  {\bibfnamefont {N.~N.}\ \bibnamefont {Lathiotakis}}, \ and\ \bibinfo {author}
  {\bibfnamefont {E.~K.~U.}\ \bibnamefont {Gross}},\ }\href
  {http://stacks.iop.org/1367-2630/17/i=9/a=093038} {\bibfield  {journal}
  {\bibinfo  {journal} {New Journal of Physics}\ }\textbf {\bibinfo {volume}
  {17}},\ \bibinfo {pages} {093038} (\bibinfo {year}
  {2015}{\natexlab{b}})}\BibitemShut {NoStop}%
\bibitem [{\citenamefont {Becke}(1993)}]{becke93_jcp98_1372}%
  \BibitemOpen
  \bibfield  {author} {\bibinfo {author} {\bibfnamefont {A.~D.}\ \bibnamefont
  {Becke}},\ }\href {\doibase 10.1063/1.464304} {\bibfield  {journal} {\bibinfo
   {journal} {J. Chem. Phys.}\ }\textbf {\bibinfo {volume} {98}},\ \bibinfo
  {pages} {1372} (\bibinfo {year} {1993})}\BibitemShut {NoStop}%
\bibitem [{\citenamefont {Savin}(1988)}]{savin88_ijqc34_59}%
  \BibitemOpen
  \bibfield  {author} {\bibinfo {author} {\bibfnamefont {A.}~\bibnamefont
  {Savin}},\ }\href {\doibase 10.1002/qua.560340811} {\bibfield  {journal}
  {\bibinfo  {journal} {Int. J. Quantum Chem. Suppl. 22}\ }\textbf {\bibinfo
  {volume} {34}},\ \bibinfo {pages} {59} (\bibinfo {year} {1988})}\BibitemShut
  {NoStop}%
\bibitem [{\citenamefont {Heyd}\ \emph {et~al.}(2003)\citenamefont {Heyd},
  \citenamefont {Scuseria},\ and\ \citenamefont
  {Ernzerhof}}]{heyd03_jcp118_8207}%
  \BibitemOpen
  \bibfield  {author} {\bibinfo {author} {\bibfnamefont {J.}~\bibnamefont
  {Heyd}}, \bibinfo {author} {\bibfnamefont {G.}~\bibnamefont {Scuseria}}, \
  and\ \bibinfo {author} {\bibfnamefont {M.}~\bibnamefont {Ernzerhof}},\ }\href
  {\doibase doi:10.1063/1.1564060} {\bibfield  {journal} {\bibinfo  {journal}
  {J. Chem. Phys.}\ }\textbf {\bibinfo {volume} {118}},\ \bibinfo {pages}
  {8207} (\bibinfo {year} {2003})}\BibitemShut {NoStop}%
\bibitem [{\citenamefont {Perdew}\ \emph {et~al.}(1982)\citenamefont {Perdew},
  \citenamefont {Parr}, \citenamefont {Levy},\ and\ \citenamefont
  {Balduz}}]{perdew82_prl49_1691}%
  \BibitemOpen
  \bibfield  {author} {\bibinfo {author} {\bibfnamefont {J.~P.}\ \bibnamefont
  {Perdew}}, \bibinfo {author} {\bibfnamefont {R.~G.}\ \bibnamefont {Parr}},
  \bibinfo {author} {\bibfnamefont {M.}~\bibnamefont {Levy}}, \ and\ \bibinfo
  {author} {\bibfnamefont {J.~L.}\ \bibnamefont {Balduz}},\ }\href {\doibase
  doi:10.1103/PhysRevLett.49.1691} {\bibfield  {journal} {\bibinfo  {journal}
  {Phys. Rev. Lett.}\ }\textbf {\bibinfo {volume} {49}},\ \bibinfo {pages}
  {1691} (\bibinfo {year} {1982})}\BibitemShut {NoStop}%
\bibitem [{\citenamefont {Sharma}\ \emph
  {et~al.}(2008{\natexlab{b}})\citenamefont {Sharma}, \citenamefont {Dewhurst},
  \citenamefont {Lathiotakis},\ and\ \citenamefont
  {Gross}}]{PhysRevB.78.201103}%
  \BibitemOpen
  \bibfield  {author} {\bibinfo {author} {\bibfnamefont {S.}~\bibnamefont
  {Sharma}}, \bibinfo {author} {\bibfnamefont {J.~K.}\ \bibnamefont
  {Dewhurst}}, \bibinfo {author} {\bibfnamefont {N.~N.}\ \bibnamefont
  {Lathiotakis}}, \ and\ \bibinfo {author} {\bibfnamefont {E.~K.~U.}\
  \bibnamefont {Gross}},\ }\href {\doibase 10.1103/PhysRevB.78.201103}
  {\bibfield  {journal} {\bibinfo  {journal} {Phys. Rev. B}\ }\textbf {\bibinfo
  {volume} {78}},\ \bibinfo {pages} {201103} (\bibinfo {year}
  {2008}{\natexlab{b}})}\BibitemShut {NoStop}%
\bibitem [{\citenamefont {Pernal}\ and\ \citenamefont
  {Cioslowski}(2005)}]{Pernal200571}%
  \BibitemOpen
  \bibfield  {author} {\bibinfo {author} {\bibfnamefont {K.}~\bibnamefont
  {Pernal}}\ and\ \bibinfo {author} {\bibfnamefont {J.}~\bibnamefont
  {Cioslowski}},\ }\href {\doibase
  http://dx.doi.org/10.1016/j.cplett.2005.06.103} {\bibfield  {journal}
  {\bibinfo  {journal} {Chemical Physics Letters}\ }\textbf {\bibinfo {volume}
  {412}},\ \bibinfo {pages} {71 } (\bibinfo {year} {2005})}\BibitemShut
  {NoStop}%
\end{thebibliography}
%

\end{document}